\pdfoutput=1
\documentclass[aps,showpacs,twocolumn,twoside,pra,superscriptaddress]{revtex4-2}
\usepackage{epsfig,amsmath,amssymb,amsfonts}
\usepackage{graphicx}
\usepackage{array}
\usepackage{xcolor}
\usepackage[unicode]{hyperref}

\graphicspath{{figures/}}

\hypersetup{
	pdffitwindow=true,
	colorlinks=true,
	urlcolor=cyan,            
	linkcolor=red,
	citecolor=green,
	filecolor=magenta
}

\def\I#1{\!#1\!}
\def\vecb#1{\boldsymbol{#1}}
\def\ket#1{|#1\rangle}
\def\bra#1{\langle#1|}
\def\scal#1#2{\langle#1|#2\rangle}
\def\matr#1#2#3{\langle#1|#2|#3\rangle}
\def\abs#1{\left\lvert#1\right\rvert}

\def\ave#1{\langle#1\rangle}

\def\abs#1{\left|#1\right|}
\newcommand{\ui}{\mathrm{i}}

\def\uvo#1{\lq\lq #1\rq\rq}

\begin{document}

\title{Stabilization of product states and excited-state quantum phase transitions \\
in a coupled qubit-field system}

\author{Pavel Str\'ansk\'y}
\email{stransky@ipnp.mff.cuni.cz}
\affiliation{Institute of Particle and Nuclear Physics, Faculty of Mathematics and Physics, Charles University, V Hole\v{s}ovi\v{c}k\'ach 2, 18000 Prague, Czech Republic}
\author{Pavel Cejnar}
\email{cejnar@ipnp.mff.cuni.cz}
\affiliation{Institute of Particle and Nuclear Physics, Faculty of Mathematics and Physics, Charles University, V Hole\v{s}ovi\v{c}k\'ach 2, 18000 Prague, Czech Republic}
\author{Radim Filip}
\email{filip@optics.upol.cz}
\affiliation{Department of Optics, Faculty of Science, Palack{\'y} University, 17.\,listopadu 1192/12, 77146 Olomouc, Czech Republic}

\date{\today}

\begin{abstract}
We study a system of a single qubit (or a few qubits) interacting with a soft-mode bosonic field.
Considering an extended version of the Rabi model with both parity-conserving and parity-violating interactions, we disclose a complex arrangement of quantum phase transitions in the ground- and excited-state domains.
An experimentally testable signature of some of these transitions is a dynamical stabilization of a fully factorized qubit-field state involving the field vacuum. 
It happens in the ultrastrong coupling regime where the superradiant field equilibrium is far from the vacuum state. 
The degree of stabilization varies abruptly with interaction parameters and increases with the softness of the field mode. 
We analyze semiclassical origins of these effects and show their connection to various forms of excited-state quantum phase transitions.  
\end{abstract}

\maketitle

\section{Introduction}
\label{Intro}

Controlled evolution of quantum systems becomes a~major challenge of modern physics in relation to quickly advancing technology \cite{Qcomp,Geo14}.
Quantum computers and simulators are presented as devices, potentially capable to hugely surpass the efficiency and the range of use of the present-day computing facilities.
One of the biggest obstacles on the way to future realization of these machines follows from the necessity to protect the quantum state of the controlled system against the effects of decoherence induced by unmanageable interactions with the environment \cite{Har96}.
Several approaches to tackle this difficulty have been developed over the past decades \cite{Qerr}, but the problem still remains crucial.

In this paper we report on a spontaneous quantum state protection in a system of ${N\I{\gtrsim}1}$ qubits (spins or two-level atoms) strongly interacting with a scalar bosonic field.
We use an extension of the Rabi model \cite{Rab36}, a version of the Dicke atom-field model \cite{Dic54,Jay63,Hep73,Wan73,Rza75,Ema03,Bra05,Bau11,Klo17,Bas17} with a single atom (or, more generally, with a small number of atoms).
The Rabi model is a paradigm for interacting quantum systems with no classical counterpart \cite{Bra11}.
It plays an important role in the present theoretical and experimental efforts to understand and control quantum dynamics in a system which provides a possible platform for quantum information applications.
The quantum dynamics of the Rabi model has been recently investigated experimentally in superconducting circuit quantum electrodynamics \cite{For10,Nie10,Bra17,For17,Yos17,Lan17}, quantum photonics \cite{Cre12}, semiconductors \cite{Tod09,Gun09}, and trapped ions \cite{Lv18}.
In quantum information processing, the interaction Hamiltonian of the Rabi form describes the hybrid gate essential for building large and complex superposition states \cite{Lo15,Kie16,Flu18,Flu19}. 
The Rabi coupling has been proposed to build necessary ingredients for quantum technology \cite{Par17,Par18,Sta20,Has21a,Has21b,Che21,Ste19}.
A~combination of the Rabi coupling with a~parity-violating interaction has been used to predict a new regime of multi-level excitations \cite{Gar16,Gar20}, reach nonlinear couplings between the qubits \cite{Koc17a,Koc17b,Sta17} and obtain quantum coherences autonomously from a cold bath \cite{Gua18,Pur20}. 
The Rabi model has been also extensively studied in connection with various new types of quantum criticality \cite{Ash13,Hwa15,Pue16a,Pue16,Pue18,Pue20,Pue20b}.
Recent experimental verification of its ground-state quantum phase transition \cite{Cai21} is very likely to stimulate serious efforts to detect also the other quantum critical effects.

In the present theoretical study, the Rabi model with both parity-conserving and parity-violating interactions is employed to address the problem of durability of separated qubit states in a strongly interacting environment.
We initiate the system at zero temperature in absence of qubit-field interaction, in a product state, in which the qubits are in a pure state $\ket{\psi}^{\otimes N}$ (carrying replicas of an arbitrary one-qubit state $\ket{\psi}$) and the boson field is in the vacuum state $\ket{0}$.
This entirely separable state is shown to exhibit a surprisingly large degree of stabilization even if a strong qubit-field interaction alters the equilibrium state of the field from the vacuum to the superradiant form.
The degree of stabilization increases with decreasing energy of an elementary field excitation and manifests abrupt variations with parameters of the qubit-field interaction.
A stronger interaction in some cases leads to even stronger stabilization.
We show that these properties result from the presence of various types of excited-state quantum phase transitions (ESQPTs) in the energy spectrum.

The ESQPTs represent an extension of the ground-state quantum phase transitions (QPTs) \cite{Car10} to the excited domain.
They attract a lot of recent attention in the context of diverse interacting quantum many-body systems, see, e.g., 
Refs.\,\cite{Cej06,Cap08,Cej08,Per09,Per11,Lar13,Bra13,Die13,Bas14a,Bas14,San15,San16,Str16,Rel16,Klo18,Kha19,Mac19,Wan19,Hum19,Tia20,Mum20,Str20,Fel21,Klo21,Cab21} and the review \cite{Cej21}.
The ESQPTs in their most common incarnations reflect the presence of saddle points or local extremes in energy landscapes associated with the classical-limit Hamiltonians, hence they generically cope with spontaneous stabilization effects like that outlined above.
Indeed, for the present model we show that although the normal vacuum becomes unstable in the superradiant phase, it still represents an excited quasistationary point in some parameter domains. 
Stability properties of dynamics near this point change with the model parameters, which leads to ESQPTs of different types and to the different degrees of the initial-state stabilization.
The effect is endemic in Rabi systems with parity-conserving interactions, but exists also in presence of some parity-violating interactions that reshape quantum critical properties and introduce new coherence phenomena.
These results are of fundamental interest as they help to identify various experimentally detectable dynamical signatures of criticality in small quantum systems. 
Moreover, they form physical background for possible future realizations of quantum information procedures based on such systems.

The plan of the paper is as follows:
In Sec.\,\ref{Model} we introduce the extended Rabi model and in Sec.\,\ref{Classical} we analyze its classical limit and quantum critical properties.
We show that the model in its extended form has a rather rich quantum phase structure with various types of QPTs and ESQPTs.
In Sec.\,\ref{Dynamics} we describe the quantum dynamics of a product qubit-field initial state and discuss the conditions needed for the stabilization effect to be observed.
The occurrence and intensity of the effect depends on the types of ESQPTs and their configuration in the energy spectrum.
In Sec.\,\ref{Clouseau} we formulate conclusions.

\section{Extended Rabi model}
\label{Model}

We consider an ensemble of identical qubits interacting with a scalar single-mode boson field.
The number $N$ of qubits will be mostly set to ${N=1}$, in which case our model represents a generalization of the Rabi model.
Nevertheless, our results remain relevant also in the context of a generalized Dicke model with moderate values ${N\gtrsim 1}$.
The field quanta are created and annihilated by operators~$\hat{b}^{\dag}$ and~$\hat{b}$ and the Hilbert space ${\cal H}_{b}$ of the field subsystem is spanned by the basis $\{\ket{n}_{b}\}_{n=0}^{\infty}$ of all eigenvectors of the boson number operator ${\hat{n}=\hat{b}^{\dag}\hat{b}}$.
The qubits (e.g., spin-$\frac{1}{2}$ particles or two-level atoms) are described by the collective quasispin operators ${\hat{\vecb{J}}\equiv(\hat{J}_x,\hat{J}_y,\hat{J}_z)}$,
\begin{equation}
\hat{\vecb{J}}=\frac{1}{2}\sum_{i=1}^{N}\hat{\vecb{\sigma}}^{(i)},
\label{quas}
\end{equation}
where ${\hat{\vecb{\sigma}}^{(i)}\equiv(\hat{\sigma}^{(i)}_x,\hat{\sigma}^{(i)}_y,\hat{\sigma}^{(i)}_z)}$ are Pauli matrices acting in the two-dimensional Hilbert space of the $i$th qubit and the components of $\hat{\vecb{J}}$ satisfy the commutation relations of angular momentum.
In the present model, the squared quasispin $\hat{\vecb{J}}^2$ is always conserved and we will assume that its quantum number $j$ takes the maximal value 
\begin{equation}
j=\frac{1}{2}\,N,
\end{equation} 
so the qubit dynamics is restricted into the exchange-symmetric ${(2j\I{+}1)}$-dimensional space ${\cal H}_{q}$ spanned by the basis $\{\ket{m}_q\}_{m=-j}^{+j}$ of the eigenvectors of the operator~$\hat{J}_z$.
In general, this is a subspace of the entire $2^N$-dimen\-sional qubit Hilbert space spanned by all combinations of the logical states of individual qubits. 
The Hilbert space of the entire qubit-field system is a tensor product ${\cal H}={\cal H}_{q}\otimes{\cal H}_{b}$, with a possible set of basis vectors ${\ket{m}_{q}\otimes\ket{n}_{b}\equiv\ket{m,\!n}}$.

In absence of interaction, each bosonic quantum has an energy $\omega$ and the energies assigned to the $\pm\frac{1}{2}$ spin projections of each qubit (the eigenstates of $\hat{\sigma}^{(i)}_z$) are $\pm\frac{1}{2}\omega_{q}$ ({we choose ${\omega,\omega_{q}>0}$).
In the following, the ratio $R=\omega_{q}/\omega$ of the single-qubit and single-boson energies will be assumed to be very large, so $R\gg 1$.
The Hamiltonian of the whole qubit-field system is expressed as
\begin{equation}
\hat{H}=\hat{H}_0(\omega,R)+\hat{H}_{\rm int}(\lambda,\delta,\mu,\gamma),
\label{Ham}
\end{equation}
where the first term represents the free Hamiltonian of both the field and qubit subsystems,
\begin{equation}
\hat{H}_0(\omega,R)=\omega\,\left[\hat{b}^{\dag}\hat{b}+R\,\hat{J}_z\right],
\label{H0}
\end{equation}
and the second term describes an interaction of these subsystems.
The interaction in the most general form considered here has the form
\begin{eqnarray}
\frac{\hat{H}_{\rm int}(\lambda,\delta,\mu,\gamma)}{2\sqrt{NR}}&=&\lambda
\overbrace{\left[\left(\hat{b}^{\dag}\I{+}\hat{b}\right)\hat{J}_x-i\delta\left(\hat{b}^{\dag}\I{-}\hat{b}\right)\hat{J}_y\right]}^{\frac{1+\delta}{2}\left(\hat{b}^{\dag}\hat{J}_{-}+\hat{b}\hat{J}_{+}\right)+\frac{1-\delta}{2}\left(\hat{b}^{\dag}\hat{J}_{+}+\hat{b}\hat{J}_{-}\right)}
\nonumber\\
&+&\ \mu\,\left(\hat{b}^{\dag}\I{+}\hat{b}\right)\left(\hat{J}_z\I{+}\gamma\,j\right),
\label{Hint}
\end{eqnarray}
with ${\hat{J}_{\pm}=\hat{J}_{x}\pm\hat{J}_{y}}$ standing for the standard angular-momentum ladder operators. 
The interaction Hamiltonian is controlled by four adjustable parameters, namely ${\lambda,\mu\in[0,\infty)}$, ${\delta\in[-1,+1]}$ and ${\gamma=0}$ or 1.
The meaning of these parameters and the $\sqrt{NR}$ scaling in Eq.\,\eqref{Hint} will be explained below.

We first discuss the Hamiltonian with ${\mu=0}$.
In this case, the model represents an ${N=1}$ (or ${N\gtrsim 1}$) analog of the extended Dicke Hamiltonian in the form considered, e.g., in Ref.\,\cite{Klo17}.
While the case with ${\delta\in[0,1]}$ was studied in the extended Dicke setup, the case with ${\delta\in[-1,0)}$ is new in the present parametrization.
The following special forms of the ${\mu=0}$ Rabi Hamiltonian can be distinguished: (i) Dicke-like form with ${\delta=0}$, (ii) Jaynes-Cummings form with ${\delta=+1}$, and (iii) anti-Jaynes-Cummings form with ${\delta=-1}$.
The Dicke-like form coincides with the original version of the Dicke Hamiltonian \cite{Dic54}, with both normal and counter-rotating terms ${(\hat{b}^{\dag}\hat{J}_{-}+\hat{b}\hat{J}_{+})}$ and ${(\hat{b}^{\dag}\hat{J}_{+}+\hat{b}\hat{J}_{-})}$ equally weighted.
On the other hand, the Jaynes-Cummings \cite{Jay63} and anti-Jaynes-Cummings forms represent two opposite extremes of simplification which completely disregard the counter-rotating or the normal term, respectively. 

We observe that any ${\mu=0}$ Hamiltonian conserves a generalized parity
\begin{equation}
\hat{\Pi}=(-1)^{\hat{n}}(-1)^{\hat{n}_{*}},
\label{parit}
\end{equation}
where $\hat{n}$ is the number of bosons and ${\hat{n}_{*}=\hat{J}_z+j}$ the number of spin-up (excited) qubits.
In addition, the Hamiltonian with ${\delta=+1}$ or ${\delta=-1}$ conserves also the quantity $\hat{M}_{+}$ or $\hat{M}_{-}$, respectively, where
\begin{equation}
\hat{M}_{\pm}=\hat{n}\pm\hat{n}_{*}.
\end{equation}
Therefore, both the Jaynes-Cummings and anti-Jaynes-Cummings limits of the ${\mu=0}$ model are integrable in the usual sense (for a more general discussion of quantum integrability of the Rabi model see Ref.\,\cite{Bra11}).

The parity \eqref{parit} is not conserved by the Hamiltonian \eqref{Hint} with ${\mu\neq 0}$.
We will see that the parity-violating term produces qualitatively new coherence effects in the qubit dynamics even in the perturbative regime with very small values of $\mu$.
We consider two parity-violating regimes: the one with ${\gamma=1}$, showing a modified type of quantum phase transition to the superradiant ground-state phase, and the one with ${\gamma=0}$, showing only a smooth crossover behavior. 
While the ${\gamma=1}$ Hamiltonian brings the parity violation via a combination of the qubit-field coupling and an external drive of the field, in the ${\gamma=0}$ case the external drive is missing \cite{Gua18,Pur20}.
We note that the parity-violating term used in our Hamiltonian differs from the term ${\propto\hat{J}_x}$ commonly employed for the same purpose in the literature (see, e.g., Ref.\,\cite{Bra11}), but with the aid of convenient rescaling of operators and parameters
these alternative Hamiltonians can be transformed to the present form with ${\gamma=0}$.
Let us also mention a possible generalization of the parity-violating Hamiltonian by introducing an independent and continuous
control of the ${\propto j}$ term. 
This however brings no essentially new feature and is not considered here.

Critical behavior, in general, is an emergent property achieved in its sharp form only in the system's infinite-size limit.
So the thermal phase transitions \cite{Hep73,Wan73,Rza75} as well as the QPTs \cite{,Ema03} and ESQPTs \cite{Bra13,Bas14a,Klo17} in the the Dicke model can be observed only if the number of atoms ${N\to\infty}$. 
However, in the Rabi model $N$ is assumed to be small, later even set to ${N=1}$.
The possibility to observe quantum critical effects in this case relies on the increase of the qubit-to-boson energy ratio $R$ \cite{Ash13,Hwa15,Pue16}.
Ratios ${R\gg 1}$ imply that the expectation values of the number of bosons in typical eigenstates of the entire system satisfy ${\ave{\hat{n}}\gg 1}$.
On the total energy scale ${E\sim N\omega_{q}}$ (which follows from the required ability to reach the excitation of the whole qubit subsystem) we expect values of the order of magnitude ${\ave{\hat{n}}\sim NR}$.
Thus in the present case, $NR$ (or just $R$) plays the role of a suitable size parameter and ${R\to\infty}$ is identified with the infinite-size limit.
Note that a systematic definition of size parameters in imbalanced coupled systems is described in Ref.\,\cite{Cej21}.

The present form of the Hamiltonian in Eqs.\,\eqref{Ham}--\eqref{Hint} implies that for increasing $R$ the energy scale (hence also the time scale) of the field subsystem remains fixed, while the energy scale of the qubit subsystem increases (the time scale decrease).
The necessity to boundlessly increase the ratio $R$ explains the use of the $\sqrt{NR}$ scaling of the interaction Hamiltonian \eqref{Hint}.
It ensures a stable proportion between expectation values $\ave{\hat{H}_{\rm int}}$ and $\ave{\hat{H}_{0}}$ in typical eigenstates when ${R\to\infty}$ (otherwise $\ave{\hat{H}_{0}}$ would become increasingly dominant). 
Note that the scaling factor for each specific value of $R$ can be absorbed into some renormalized interaction strengths.

To facilitate the description of the system with large values of the size parameter, we transform the Hamiltonian \eqref{Ham} into the coordinate-momentum form of the boson operators.
In particular, we use the mapping
\begin{equation}
\hat{b}^{\dag}=\sqrt{\frac{NR}{2}}\left(\hat{q}-\ui\hat{p}\right),\quad 
\hat{b}=\sqrt{\frac{NR}{2}}\left(\hat{q}+\ui\hat{p}\right), 
\label{qp}
\end{equation}
where $\hat{q}$ and $\hat{p}$, respectively, represent canonical coordinate and momentum of the field subsystem, with the commutation relation ${[\hat{q},\hat{p}]=\ui/NR}$ identifying  $1/NR$ as an effective Planck constant.
We will use a scaled dimensionless Hamiltonian and the corresponding energy
\begin{equation}
\hat{h}=\frac{\hat{H}}{NR\,\omega},\qquad \varepsilon=\frac{E}{NR\,\omega}.
\end{equation}
In terms of $\hat{q}$ and $\hat{p}$, the scaled Hamiltonian reads
\begin{eqnarray}
\hat{h}=-\frac{1}{2NR}&+&\frac{\hat{q}^2\I{+}\hat{p}^2}{2}+\sqrt{2}N\,\frac{\mu\gamma}{\omega}\,\hat{q}
\label{Hpq}\\
&&+\underbrace{\left(\sqrt{8}\,\frac{\lambda}{\omega}\,\hat{q}\,,\,-\sqrt{8}\,\frac{\lambda\delta}{\omega}\,\hat{p}\,,\,\frac{1}{N}\I{+}\sqrt{8}\,\frac{\mu}{\omega}\,\hat{q}\right)}_{\hat{\vecb{B}}}\cdot\,\hat{\vecb{J}},
\nonumber
\end{eqnarray}
where we introduced a vector of operators $\hat{\vecb{B}}\equiv{(\hat{B}_x,\hat{B}_y,\hat{B}_z)}$ which enabled us to rewrite the qubit-field interaction Hamiltonian in the magnetic-dipole form ${\hat{\vecb{B}}\cdot\hat{\vecb{J}}}\equiv{\hat{B}_x\hat{J}_x\I{+}\hat{B}_y\hat{J}_y\I{+}\hat{B}_z\hat{J}_z}$.
This expression will be employed in the semiclassical analysis below.

\section{Semiclassical analysis and quantum phase transitions}
\label{Classical}

\subsection{General expressions}
\label{Clasgen}

In the limit ${R\to\infty}$ the coordinate and momentum operators $\hat{q}$ and $\hat{p}$ in Eq.\,\eqref{qp} satisfy ${[\hat{q},\hat{p}]\to 0}$, so they can be treated as ordinary commuting variables~$q$ and~$p$.
Hence the vector operator $\hat{\vecb{B}}$ in Eq.\,\eqref{Hpq} maps onto an ordinary c-number vector $\vecb{B}$ and the interaction term can be cast with the aid of the quasispin projection in a rotated $z'$ direction parallel with $\vecb{B}$, so
\begin{equation}
\vecb{B}\cdot\hat{\vecb{J}}=\sqrt{\frac{8\lambda^2}{\omega^2}\,q^2\I{+}\frac{8\lambda^2\delta^2}{\omega^2}\,p^2\I{+}\left(\frac{1}{N}\I{+}\frac{\sqrt{8}\mu}{\omega}\,q\right)^2}\,\underbrace{\vecb{n}\cdot\hat{\vecb{J}}}_{\hat{J}_{z'}},
\quad\label{BJ}\\
\end{equation}
where ${\vecb{n}=\vecb{B}/|\vecb{B}|\equiv(n_x,n_y,n_z)}$ is a unit vector which depends on~$q$ and~$p$. 
Under these circumstances, the projection $\hat{J}_{z'}$ commutes with the Hamiltonian \eqref{Hpq} and therefore represents an exact integral of motion.
Let us stress that this conservation law is not satisfied if $R$ is finite since then the unit vector $\vecb{n}$ in the definition of $\hat{J}_{z'}$ becomes an operator $\hat{\vecb{n}}$ depending on $\hat{q}$ and $\hat{p}$ which does not commute with the other parts of the Hamiltonian (in that case even the expression of $\hat{\vecb{B}}$ in terms of $|\hat{\vecb{B}}|$ and $\hat{\vecb{n}}$ would have to be performed in a~more careful way respecting the Hermicity requirement). 
If $R$ is finite but large enough, the correctly defined quasispin projection $\hat{J}_{z'}$ represents an approximate integral of motion following from an approximate adiabatic separation of fast qubit dynamics from the slow field dynamics \cite{Bas17}.
This means that the spectrum of the full Hamiltonian \eqref{Hpq} decomposes into virtually noninteracting subsets of states with different values of the $\hat{J}_{z'}$ quantum number $m'=-j,-j\I{+}1,...,j\I{-}1,+j$.

The ${R\to\infty}$ classical Hamiltonian depending on variables $q$ and $p$ for a given value $m'$ of $\hat{J}_{z'}$ follows from Eqs.\,\eqref{Hpq} and \eqref{BJ}: 
\begin{eqnarray}
h_{m'}(q,p)=&&\frac{q^2\I{+}p^2}{2}+\sqrt{2}N\,\frac{\mu\gamma}{\omega}\,q
\label{Hacla}\\
&&+m'\sqrt{\frac{8\lambda^2}{\omega^2}\,q^2\I{+}\frac{8\lambda^2\delta^2}{\omega^2}\,p^2\I{+}\left(\frac{1}{N}\I{+}\frac{\sqrt{8}\mu}{\omega}\,q\right)^2}.
\nonumber
\end{eqnarray}
This Hamiltonian defines classical dynamics of the single degree of freedom (${f=1}$) associated with the field subsystem constrained by the state of the qubit subsystem with the fixed quasispin projection $m'$.
If $R$ is finite but large enough, the discrete energy spectra of the whole system can be obtained by the semiclassical quantization of the field degree of freedom performed separately for each value of $m'$ in Eq.\,\eqref{Hacla}.
The lowest state for a given $m'$ approximately coincides with the minimum of $h_{m'}(q,p)$, while the density of excited states is obtained from the formula
\begin{equation}
\bar{\rho}_{m'}(\varepsilon)=\frac{1}{2\pi}\,\frac{d}{d\varepsilon}\int dq\,dp\ \Theta\bigl(\varepsilon\I{-}h_{m'}(q,p)\bigr),
\label{leden}
\end{equation}
where $\Theta(x)$ is the step function (${\Theta=0}$ and 1, respectively, for ${x<0}$ and ${\geq 1}$).
The bar in $\bar{\rho}_{m'}$ indicates that the semiclassical (infinite-size) level density \eqref{leden} approximates smoothed level densities in the finite-size cases.
Note the symmetry of the Hamiltonian \eqref{Hacla} under the parameter change $\delta\to-\delta$, which makes the semiclassical results discussed below dependent only on $|\delta|$.

Identification of quantum critical effects relies on the analysis of stationary points of the set of Hamiltonian functions in Eq.\,\eqref{Hacla}.
A nonanalytic change of the absolute minimum of $h_{m'}(q,p)$ at some value of the control parameter indicates a QPT of the lowest state in the given $m'$-subset.
Any stationary point of $h_{m'}(q,p)$ at energy $\varepsilon_{\rm c}$ above the global minimum causes an ESQPT nonanalyticity in the level density \eqref{leden}.
As shown in Ref.\,\cite{Str16}, for a nondegenerate (locally quadratic) stationary point with index $r$ (the number of negative eigenvalues of the Hessian matrix) the nonanalyticity has the form,
\begin{equation}
\bar{\rho}_{m'}(\varepsilon)-\bar{\rho}^{(0)}_{m'}(\varepsilon)
\propto\!\left\{\!\!\begin{array}{ll}
(-1)^{r/2}\,\Theta(\varepsilon\I{-}\varepsilon_{\rm c}) & {\rm for\ }r=0,2,\\
-\ln|\varepsilon\I{-}\varepsilon_{\rm c}| & {\rm for\ }r=1,
\end{array}\right.
\label{Esqpts}
\end{equation}
where a model-specific analytic part $\bar{\rho}^{(0)}_{m'}(\varepsilon)$ of the full level density, which depends on the behavior of the Hamiltonian away from the stationary point, is subtracted to gain the universal irregular part, which reflects only an infinitesimal vicinity of the stationary point.
As seen in Eq.\,\eqref{Esqpts}, at the energy $\varepsilon_{\rm c}$ of the nondegenerate stationary point the semiclassical level density manifests either an upward or downward step discontinuity, or a logarithmic divergence.

\subsection{Parity conserving case}
\label{Clascons}

\begin{figure}[t]
\includegraphics[width=\linewidth]{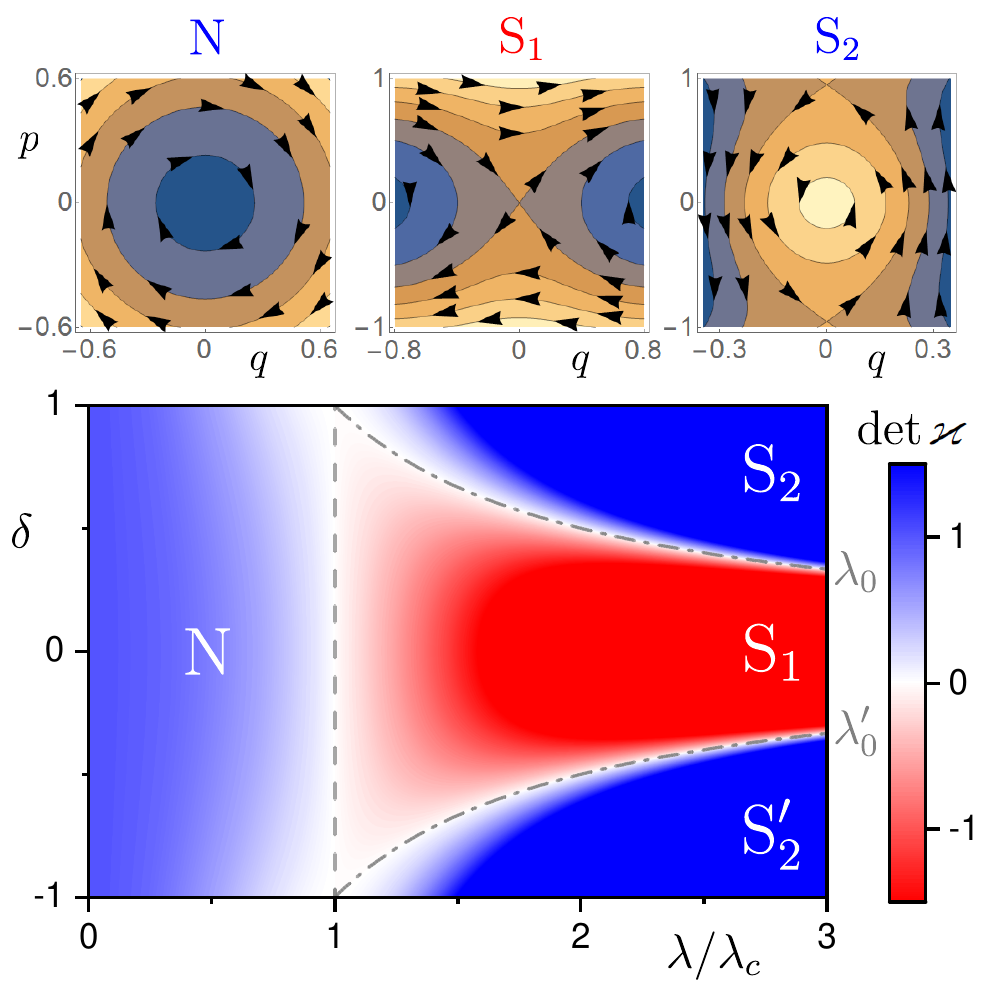}
\caption{
\uvo{Quantum phases} of the field vacuum, i.e., the types of the $(q,p)\I{=}(0,0)$ stationary point, in the parameter plane of the parity-conserving (${\mu=0}$) Hamiltonian with ${m'=-j}$.
The dashed line at ${\lambda=\lambda_{\rm c}}$ and the dash-dotted curves denoted as $\lambda_{0}$ and $\lambda_{0}'$, which represent the $\lambda_{0}(\delta)$ dependence from Eq.\,\eqref{lac} for ${\delta>0}$ and ${\delta<0}$, respectively, separate the normal phase~${\rm N}$, the first superradiant phase ${\rm S}_1$ and the second superradiant phases  ${\rm S}_2$ and ${\rm S}'_2$.
The color encodes the value of the Hessian matrix determinant at $(q,p)\I{=}(0,0)$.
The insets show dynamics of perturbations near the stationary point for some $(\lambda,\delta)$ in the ${\rm N}$, ${\rm S}_1$ and  ${\rm S}_2$ phases.
}
\label{phase}
\end{figure}

\begin{figure}[t]
\includegraphics[width=\linewidth]{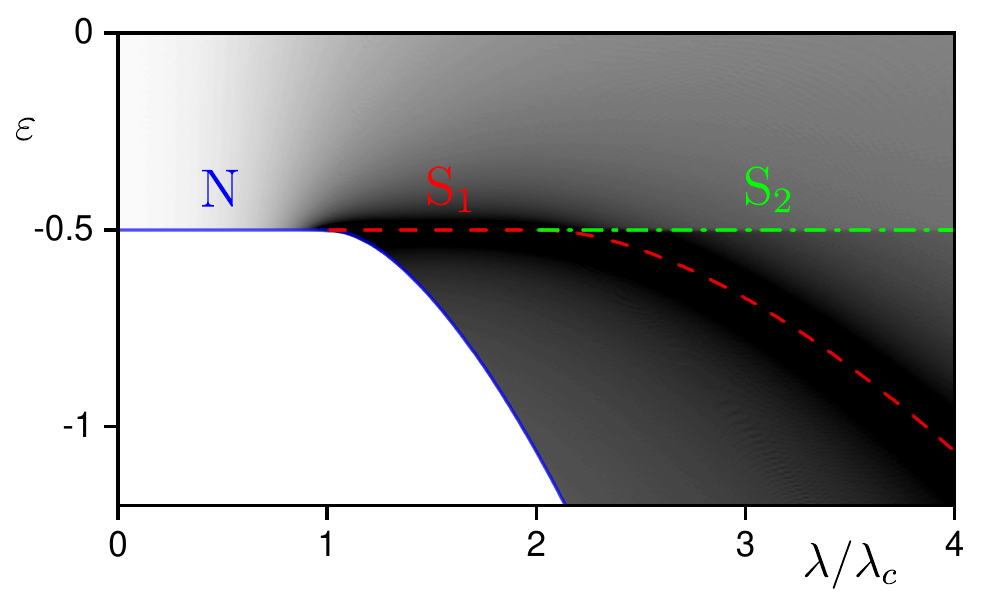}
\caption{
The semiclassical level density $\bar{\rho}_{-j}(\varepsilon)$ from Eq.\,\eqref{leden} as a function of $\varepsilon$ and $\lambda$ for the parity-conserving (${\mu=0}$) Hamiltonian~\eqref{Ham} with ${|\delta|=0.5}$ and ${\omega=1=N}$. 
Darker shades correspond to larger values of the level density.
The observed ESQPT nonanalyticies of the level density (divergence and downward step, highlighted by the dashed and dash-dotted lines, respectively) result from the stationary points in Eqs.\,\eqref{qp0}--\eqref{e2}.
The lowest state undergoes a second-order QPT at ${\lambda=\lambda_{\rm c}}$.
The vacuum phases from Fig.\,\ref{phase} are indicated.
}
\label{stacpoints1}
\end{figure}

We first consider the parity-conserving ${\mu=0}$ Rabi Hamiltonians. 
Stationary points of the corresponding classical limit \eqref{Hacla} can be obtained in an analytic form for any value of $m'$.
For ${m'>0}$, the function $h_{m'}(q,p)$ has a single stationary point, the global minimum at ${(q,p)=(0,0)}$, which means that the lowest energy state is always associated with the field vacuum. 
The ${m'<0}$ cases show more interesting behavior.
In the following, we will analyze the subset of states with the lowest quasispin projection ${m'=-j}$, which represents (for ${\lambda\geq 0}$) the subspace containing the ground state.

We define two special values of parameter $\lambda$,
\begin{equation}
\lambda_{\rm c}=\frac{\omega}{2N},\qquad 
\lambda_{0}(\delta)=\frac{\lambda_{\rm c}}{|\delta|},
\label{lac}
\end{equation}
that separate, in the plane ${\lambda\times\delta}$, different \uvo{quantum phases} distinguished (as explained below) by the nature of the field vacuum in the lowest-$m'$ subspace of the entire Hilbert space, see the diagram in Fig.\,\ref{phase}.
Stationary points of $h_{-j}(q,p)$ and the corresponding QPT and ESQPT nonanalyticities of the semiclassical level density $\bar{\rho}_{-j}(\varepsilon)$ (shown for ${\delta=\pm 0.5}$ in Fig.\,\ref{stacpoints1}) can be described with reference to the above phase diagram: 

(a) For ${\lambda\in[0,\lambda_{\rm c}]}$ (the {\em normal phase\/} of the system, see the region N in Fig.\,\ref{phase}), the only stationary point is the global minimum at the position and energy
\begin{eqnarray}
(q,p)&=&(0,0)
\label{qp0}\\
\varepsilon&=&-\frac{1}{2}
\label{e0}.
\end{eqnarray}
So the field vacuum associated with the point \eqref{qp0} constitutes (along with the ${m'=m=-j}$ state of qubits) the ground state of the system, which results in the ground-state expectation values $\ave{\hat{b}^{\dag}}_{\rm gs}\I{=}\ave{\hat{b}}_{\rm gs}\I{=}\ave{\hat{b}^{\dag}\hat{b}}_{\rm gs}\I{=}0$.
The stationary point \eqref{qp0} is nondegenerate for ${\lambda<\lambda_{\rm c}}$, but it becomes degenerate (quartic) at ${\lambda=\lambda_{\rm c}}$, undergoing a pitchfork bifurcation in the $q$-direction.
This is a critical point of a second-order ground-state QPT.

(b) For ${\lambda\in(\lambda_{\rm c},\lambda_{0}]}$ (we call this domain the {\em first superradiant phase}, see the region ${\rm S}_1$ in Fig.\,\ref{phase}) we have three stationary points:
Two of them coincide with a pair of quadratic minima at
\begin{eqnarray}
(q,p)&=&\left(\pm\sqrt{\frac{1}{2}\biggl[\frac{\lambda^2}{\lambda_{\rm c}^2}-\frac{\lambda_{\rm c}^2}{\lambda^2}\biggr]},0\right),
\label{qp1}\\
\varepsilon&=&-\frac{1}{4}\biggl(\frac{\lambda^2}{\lambda_{\rm c}^2}+\frac{\lambda_{\rm c}^2}{\lambda^2}\biggr).
\label{e1}
\end{eqnarray}
These form  a doubly-degenerate ground state, on the quantum level a doublet of states with opposite values of parity \eqref{parit}.
The minima \eqref{qp1} represent new Higgs-like \uvo{vacua} of the field characterized by a nonvanishing order parameter $\ave{\hat{b}+\hat{b}^{\dag}}_{\rm gs}$, while the extremal quasispin projection $m'$ of the qubit subsystem is rotated from the original $z$-axis to a corresponding new direction.
The original field vacuum given by Eqs.\,\eqref{qp0} and \eqref{e0} is associated with the third stationary point, which is a saddle (maximum in $q$-direction, minimum in $p$-direction, index ${r=1}$).
This remnant of the ${\lambda<\lambda_{\rm c}}$ minimum (in the ${\lambda\times\varepsilon}$ plane a straight continuation of the normal-phase ground state) corresponds to an ESQPT given by a logarithmic divergence of the level density, as seen Fig.\,\ref{stacpoints1}.
At ${\lambda=\lambda_{0}}$ this saddle becomes degenerate and undergoes a pitchfork bifurcation in the $p$-direction.

(c) For ${\lambda\in(\lambda_{0},\infty)}$ (the {\em second superradiant phase}, which exists in two disconnected regions below ${\delta=+1}$ and above ${\delta=-1}$, see regions ${\rm S}_2$ and ${\rm S}'_2$ in Fig.\,\ref{phase}) there exist five stationary points.
The ground state is still given by the pair of symmetric global minima from Eqs.\,\eqref{qp1} and \eqref{e1}, while the remaining three stationary points give rise to two ESQPTs.
A degenerate pair of quadratic saddles (maxima in $q$-, minima in $p$-direction, ${r=1}$) at
\begin{eqnarray}
(q,p)&=&\left(\pm\sqrt{\frac{1}{2}\biggl[\frac{\lambda^2}{\lambda_{0}^2}-\frac{\lambda_{0}^2}{\lambda^2}\biggr]},0\right),
\label{qp2}\\
\varepsilon&=&-\frac{1}{4}\biggl(\frac{\lambda^2}{\lambda_{0}^2}+\frac{\lambda_{0}^2}{\lambda^2}\biggr),
\label{e2}
\end{eqnarray}
produces a logarithmic divergence of the level density.
The original field vacuum from Eqs.\,\eqref{qp0} and \eqref{e0} now corresponds to a local maximum in both $q$- and $p$-directions (${r=2}$), which is a remnant of the ${\lambda\in(\lambda_{\rm c},\lambda_{0}]}$ saddle.
It generates a downward step in the level density along a line continuing the ${\lambda\leq\lambda_{\rm c}}$ ground state. 
These structures are apparent in Fig.\,\ref{stacpoints1}.

As we saw, in the infinite-size limit the real vacuum of the bosonic field represents a stationary state of the system  for all ${\lambda\in[0,\infty)}$, but in the subspace ${m'=-j}$ its nature changes with increasing $\lambda$.
Indeed, the corresponding stationary point \eqref{qp0} transforms from the global minimum to a saddle point at ${\lambda=\lambda_{\rm c}}$, and then to a local maximum at ${\lambda=\lambda_{0}}$.
The Hessian matrix $\varkappa(q,p)$ of the Hamiltonian second derivatives ${\varkappa_{qq}\I{=}\frac{\partial^2}{\partial q^2}h_{-j}}$, ${\varkappa_{pp}\I{=}\frac{\partial^2}{\partial p^2}h_{-j}}$ and ${\varkappa_{qp}\I{=}\frac{\partial^2}{\partial q\partial p}h_{-j}\I{=}\frac{\partial^2}{\partial p\partial q}h_{-j}\I{=}\varkappa_{pq}}$ at $(q,p)\I{=}(0,0)$ reads as
\begin{equation}
\left(\begin{array}{cc}
\varkappa_{qq}(0,0), & \varkappa_{qp}(0,0) \\
\varkappa_{pq}(0,0), & \varkappa_{pp}(0,0)
\end{array}\right)
=
\left(\begin{array}{cc}
1\I{-}\frac{\lambda^2}{\lambda_{\rm c}^2}, 
& 
0
\\
0,
& 
1\I{-}\frac{\lambda^2}{\lambda_{0}^2}
\end{array}\right).
\label{Hesse1}
\end{equation}
The determinant of $\varkappa(0,0)$ (positive for the maximum or minimum and negative for the saddle point) is expressed by the intensity and color of the background filling in the phase diagram of Fig.\,\ref{phase}.

The dynamics of small deviations from the vacuum state \eqref{qp0} is governed by the matrix equation
\begin{equation}
\left(\begin{array}{c}
\dot{q} \\ \dot{p}
\end{array}\right)\approx
\left(\begin{array}{cc}
\varkappa_{qp}(0,0) & \varkappa_{pp}(0,0) \\
-\varkappa_{qq}(0,0) & -\varkappa_{pq}(0,0)
\end{array}\right)
\left(\begin{array}{c}
q \\ p
\end{array}\right),
\label{stabil}
\end{equation}
where we can further substitute $\varkappa_{qp}(0,0)\I{=}\varkappa_{pq}(0,0)\I{=}0$ from Eq.\,\eqref{Hesse1}.
It follows that the types of the $(0,0)$ stationary point in the three domains of $\lambda$ differ by the stability properties of nearby classical orbits:
(a) For ${\lambda\in[0,\lambda_{\rm c})}$, in the normal phase, we get ${\varkappa_{qq}>0}$ and ${\varkappa_{pp}>0}$, so the matrix in Eq.\,\eqref{stabil} has pure imaginary eigenvalues.
Hence the dynamics of deviations $(q,p)$ with ${\delta\varepsilon\equiv\varepsilon-h_{-j}(0,0)>0}$ is stable and the Hamiltonian minimum is trivially recognized as an elliptic fixed point.
(b) For ${\lambda\in(\lambda_{\rm c},\lambda_0)}$, in the first superradiant phase, we have ${\varkappa_{qq}<0}$ and ${\varkappa_{pp}>0}$, so the matrix in Eq.\,\eqref{stabil} has real eigenvalues of both signs.
The dynamics of deviations for ${\delta\varepsilon>0}$ and ${<0}$ is unstable and the saddle point of the Hamiltonian function is classified as an hyperbolic fixed point.
(c) Finally, for ${\lambda\in(\lambda_0,\infty)}$, in the second superradiant phase, we obtain ${\varkappa_{qq}<0}$ and ${\varkappa_{pp}<0}$, which again leads to pure imaginary eigenvalues of the matrix in Eq.\,\eqref{stabil}.
Thus the dynamics of deviations with $\delta\varepsilon\I{<}0$ becomes stable, the local maximum of the Hamiltonian function being identified with an elliptic fixed point.
These conclusions are illustrated in Fig.\,\ref{phase}.

We note that the form of the stationary point for ${\lambda>\lambda_0}$ (local maximum in both coordinate and momentum) is rather unusual in the context of ordinary systems of classical mechanics, but appears generically in classical limits of interacting many-body systems (see, e.g., Refs.\,\cite{Klo17,Mac19}). 
The locally stabilizing character of this stationary point will play an important role later. 

\subsection{Parity violating cases}
\label{Clasviol}

\begin{figure}[ht]
\includegraphics[width=\linewidth]{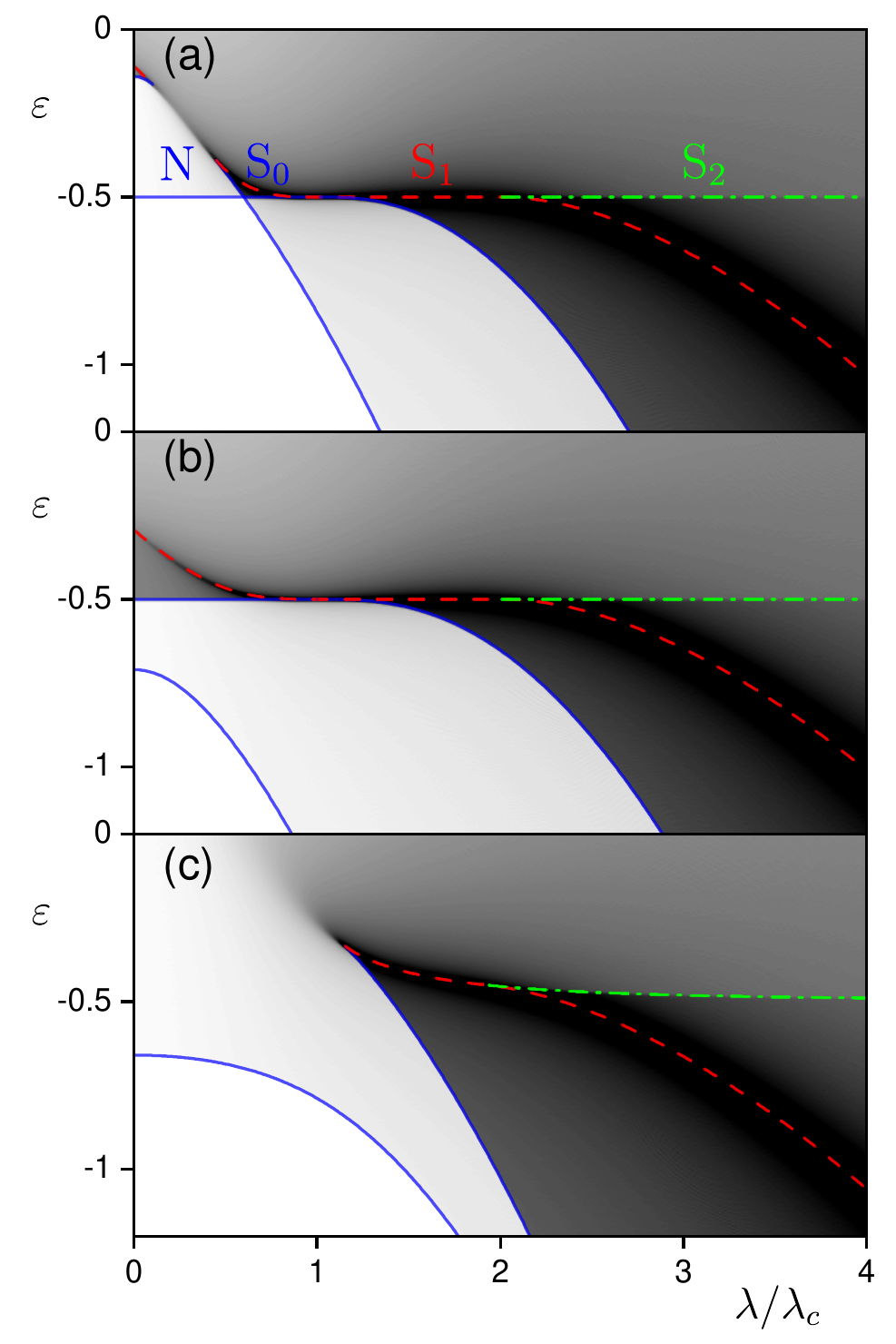}
\caption{The semiclassical level density $\bar{\rho}_{-j}(\varepsilon)$ (darker shades correspond to larger values) for parity-nonconserving versions of the Hamiltonian~\eqref{Ham} with ${|\delta|=0.5}$ and ${(\mu,\gamma)}$ taking values (a) ${(0.4,1)}$, (b) ${(0.55,1)}$, and (c) ${(0.4,0)}$.
We also set ${\omega=1=N}$.
The ESQPT nonanalyticities of the level density are highlighted by full (upwward step), dashed (divergence) and dash-dotted (downward step) lines. 
In panel (a), the ground state undergoes a fist-order QPT and the vacuum phases are indicated.}
\label{stacpoints2}
\end{figure}

Now we briefly turn to the parity violating Hamiltonians from Eqs.\,\eqref{Ham}--\eqref{Hint} with ${\mu\neq 0}$.
In these cases, the stationary-point analysis of the corresponding classical limit \eqref{Hacla} yields higher order polynomial equations and in general needs to be performed numerically.
We will consider parameters $\mu$ (as well as $\gamma$ and $\delta$) to be fixed at some constant values and analyze properties of the system as a function of the only variable parameter~$\lambda$. 
Examples of the ${m'=-j}$ semiclassical level density for three choices of  parameters $\mu$ and $\gamma$ are shown in Fig.\,\ref{stacpoints2}.

For systems with ${\mu\neq 0}$ the field vacuum may not represent any kind of equilibrium.
The condition necessary for the vacuum to be a stationary point is that ${\gamma=1}$. 
Then the equations ${\frac{\partial}{\partial q}h_{-j}\I{=}\frac{\partial}{\partial p}h_{-j}=0}$ with the Hamiltonian function \eqref{Hacla} are fulfilled at $(q,p)\I{=}(0,0)$ for any value of the other parameters.
It however turns out that the $(0,0)$ stationary point is not the global minimum for ${\mu>0.5}$ .
If both conditions ${\gamma=1}$ and ${\mu\in(0,0.5)}$ are satisfied simultaneously, the Hamiltonian driven by $\lambda$ exhibits a ground-state QPT connected with the swapping of two minima at a certain critical parameter value ${\lambda=\lambda_{\rm c}'(\mu)\in(0,\lambda_{\rm c})}$.
For ${\lambda<\lambda_{\rm c}'}$ the ground state corresponds to the field vacuum $(q,p)\I{=}(0,0)$, while for ${\lambda>\lambda_{\rm c}'}$ it is associated with the non-vacuum solution given by the lower of the two swapped minima located at a point $(q,p)\I{\neq}(0,0)$.
Therefore, the QPT at ${\lambda=\lambda_{\rm c}'(\mu)}$ for ${\mu>0}$ again represents a kind of the superradiant phase transition, but in contrast to the parity-conserving case it is of the first order, with no spontaneous symmetry breaking involved.
The critical value decreases from ${\lambda_{\rm c}'(\mu=0)=\lambda_{\rm c}}$, see Eq.\,\eqref{lac}, to ${\lambda_{\rm c}'(\mu=0.5)=0}$.

The Hessian matrix of the ${\gamma=1}$ Hamiltonian at $(q,p)\I{=}(0,0)$ is still expressed by Eq.\,\eqref{Hesse1}, which means that the three vacuum phases described in Sec.\,\ref{Clascons} exist also for ${\mu\in(0,0.5)}$.
Besides the normal phase, now found at ${\lambda\in[0,\lambda_{\rm c}')}$, and the first and second superradiant phases at ${\lambda\in(\lambda_{\rm c},\lambda_0)}$ and $(\lambda_0,\infty)$, respectively, we also observe an additional superradiant phase in the interval ${\lambda\in(\lambda_{\rm c}',\lambda_{\rm c})}$.
In this phase (which we will call the {\em zeroth superradiant phase}, ${\rm S}_0$) the vacuum state $(q,p)\I{=}(0,0)$ is still a local minimum of the Hamiltonian, though not the one associated with the lowest energy state.
The stability properties of this vacuum type are qualitatively similar to those in the normal phase. 
The QPT and ESQPT structures in the level density for a single choice of ${\mu\in(0,0.5})$  are seen in Fig.\,\ref{stacpoints2}(a).

The model with ${\gamma=1}$ and ${\mu\geq 0.5}$ does not have the vacuum ${(q,p)=(0,0)}$ as the ground state even for ${\lambda=0}$, but the vacuum still represents a stationary point above the ground state.
This is seen in Fig.\,\ref{stacpoints2}(b).
In contrast, for ${\gamma\neq 1}$ the field vacuum is no more a stationary point of any kind and plays no privileged role in the spectrum and dynamics, see Fig.\,\ref{stacpoints2}(c).
In both these parity-violating cases, the ground state eigenvector and eigenenergy evolve continuously with increasing parameter $\lambda$, with no QPT criticality taking place at any point.
Several ESQPT structures are nevertheless observed even in these cases.
If considered just in the $\lambda\times\varepsilon$ plane, they make an explicit example of \uvo{ESQPTs without a~QPT} (in the sense of Ref.\,\cite{Rel16}).
However, connections of these structures to the first- and second-order QPTs would be established in the full space of all control parameters.

\section{Quantum dynamics}
\label{Dynamics}

\subsection{General expressions}
\label{Dyngen}

We will study the evolution of the qubit-field system initially prepared in the fully noninteracting ground state at ${\lambda=\mu=0}$ by the Hamiltonian $\hat{H}$ with generally nonzero values of the interaction parameters.
Hence
\begin{equation}
\ket{\Psi(t)}=e^{-\ui\hat{H}t}\!\underbrace{\ket{m\I{=}-j}_{q}\I{\otimes}\ket{n\I{=}0}_{b}}_{\ket{\Psi(0)}}=\sum_{m,n}c_{mn}(t)\ket{m,\! n},
\label{evo}
\end{equation}
where $c_{mn}(t)$ are time-dependent coefficients of the expansion of the evolving state $\ket{\Psi(t)}$ in the basis $\ket{m,\!n}$.
Note that we set ${\hbar=1}$.
The initial state $\ket{\Psi(0)}$ is a product of the qubit and field states.
At ${t=0}$, all qubits are set to the spin-down state, which can be an arbitrary one-qubit state $\ket{\psi}\I{=}{\alpha\ket{0}\I{+}\beta\ket{1}}$ expressed in a suitable basis ($\alpha$ and $\beta$ are arbitrary amplitudes of the qubit logical states~0 and~1). 
So $\ket{m\I{=}-j}_q\I{\equiv}{\ket{\psi}^{\otimes N}}$.
The field is initiated in the state $\ket{n\I{=}0}_b$, which is the normal vacuum with no bosonic excitations present.
The fate of this simple (in fact the simplest, entirely factorized) state will depend on the choice of the evolving Hamiltonian $\hat{H}$.

It is important to realize that in the ${R\to\infty}$ limit, when the qubit and field subsystems become adiabatically separated, the evolution of the whole system receives both classical and quantum features.
While the state of the field is determined by variables $q(t)$ and $p(t)$ that evolve according to classical dynamical equations, the state of the qubits remains entirely quantal, resulting from an \uvo{external} driving performed by the time-dependent vector $\vecb{B}(t)$, which is determined by the classical field state.
Below we will study the dynamics for finite but large values of $R$.

The evolution \eqref{evo} can be seen as a response of the noninteracting system at zero temperature to the quantum quench into the interacting phase characterized by a chosen set of parameters ${(\lambda,\delta,\mu,\gamma)}$ of a general Hamiltonian from Eqs.\,\eqref{Ham}--\eqref{Hint} \cite{San15,San16,Klo18}.
The time dependence of the survival probability $P(t)$ of the initial state can be determined from the energy distribution
\begin{equation}
S(E)=\sum_i\underbrace{\abs{\scal{E_i}{\Psi(0)}}^2}_{p_i}\delta(E\I{-}E_i)
\label{strength}
\end{equation} 
(called local density of states or strength function), where $p_i$ is the probability to identify the initial state with the $i$th eigenvector $\ket{E_i}$ of $\hat{H}$.
In particular, we have 
\begin{equation}
P(t)=\left|\scal{\Psi(0)}{\Psi(t)}\right|^2=\left|\int dE\,S(E) e^{-\ui Et}\right|^2.
\label{surv}
\end{equation}
Quantum quench dynamics usually involves several stages which can be derived from the energy-time uncertainty relation $\Delta E\,\Delta t\gtrsim\tfrac{1}{2}$.  
The transient stages correspond to the time periods $\Delta t$ in which the energy uncertainty $\Delta E$ is not yet small enough to resolve all details of the distribution $S(E)$.
The evolution at very short times depends only on the overall width of the distribution, reflecting the variance ${\ave{E^2}\I{-}\ave{E}^2}\I{=}{\matr{\Psi(0)}{\hat{H}_{\rm int}^2}{\Psi(0)}}$ that grows quadratically with the overall strength of the interaction Hamiltonian~\eqref{Hint}.
On the other hand, for very long times (above the Heisenberg time scale) the full energy resolution is reached (with $\Delta E$ less than the smallest energy spacing between the populated states) and the system enters the dynamical equilibrium in which ongoing fluctuations of all quantities attain constant statistical measures.

Besides the survival probability \eqref{surv}, we consider also the time-dependent expectation values 
\begin{equation}
A(t)=\matr{\Psi(t)}{\hat{A}}{\Psi(t)},
\label{aver}
\end{equation}
where $\hat{A}$ stands for a general observable, below identified with the boson number $\hat{n}$, its quadratures $\hat{q}$ and $\hat{p}$, and with the quasispin projections $\hat{J}_x$, $\hat{J}_y$, $\hat{J}_z$.
We will study the infinite-time averages
\begin{equation}
\overline{F}=\lim_{T\to\infty}\frac{1}{T}\int_{0}^{T}dt\,F(t),
\label{avera}
\end{equation} 
where $F(t)$ coincides with the time dependencies \eqref{surv} or \eqref{aver}. 
For the averaged survival probability we get
\begin{equation}
\overline{P}=\sum_ip_i^2,
\label{asurv}
\end{equation}
which is the inverse participation ratio characterizing the initial-state distribution in the eigenbasis $\ket{E_i}$, while for the averaged expectation values of observables we obtain
\begin{equation}
\overline{A}=\sum_ip_iA_{ii},
\label{aaver}
\end{equation}
where ${A_{ii'}=\matr{E_i}{\hat{A}}{E_{i'}}}$. 

Both the qubit and field subsystems at any time $t$ can be characterized by their respective reduced density operator $\hat{\varrho}_{q}(t)$ and $\hat{\varrho}_{b}(t)$.
These are obtained by partial tracing of the density operator $\hat{\varrho}(t)\I{=}\ket{\Psi(t)}\bra{\Psi(t)}$ of the whole system over the Hilbert space of the nonparticipating subsystem:
\begin{eqnarray}
\hat{\varrho}_{q}(t)&\I{=}&{\rm Tr}_{b}\hat{\varrho}(t)\I{=}
\sum_{m,m'}\biggl[\sum_n c_{mn}(t)c^*_{m'n}(t)\biggr] \ket{m}_q\bra{m'},
\quad
\label{rhoq}\\
\hat{\varrho}_{b}(t)&\I{=}&{\rm Tr}_{q}\hat{\varrho}(t)\I{=}
\sum_{n,n'}\biggl[\sum_m c_{mn}(t)c^*_{mn'}(t)\biggr] \ket{n}_b\bra{n'}.
\label{rhob}
\end{eqnarray}
As the total state $\hat{\varrho}(t)$ is pure, both reduced density operators \eqref{rhoq} and \eqref{rhob} have the same eigenvalues, yielding therefore equal measures of the mutual entanglement.

The diagonal matrix element of the qubit density operator \eqref{rhoq} in the qubit state with ${m\I{=}-j}$ coincides with the probability to find the qubits in their initial state $\ket{\psi}^{\otimes N}$.
Hence we define the survival probability of the qubit initial state alone, 
\begin{equation}
P_q(t)=\matr{m=-j}{\hat{\varrho}_{q}(t)}{m=-j},
\label{survq}
\end{equation}
which differs from the overall survival probability in Eq.\,\eqref{surv} by disregarding the state of the field.
Similarly, the survival probability of the field state alone is expressed as the diagonal matrix element of the density operator \eqref{rhob} in the field vacuum state:
\begin{equation}
P_b(t)=\matr{n=0}{\hat{\varrho}_{b}(t)}{n=0}.
\label{survb}
\end{equation}
For fully factorized qubit-field states, i.e., when $\hat{\varrho}_{q}(t)$ and $\hat{\varrho}_{b}(t)$ represent pure states of the respective subsystems, the overall survival probability \eqref{surv} factorizes into the product of probabilities \eqref{survq} and \eqref{survb}.
This happens in the ${R\to\infty}$ limit, when the adiabatic separation of the qubit and field subsystems becomes exact, but in common finite-$R$ situations we expect partial entanglement of both subsystems and ${P(t)\neq P_q(t)P_b(t)}$.

Below we will present a numerical study of the model dynamics with a single qubit, ${N=1}$.
The reduced density operator of the qubit is expressed as a ${2\times 2}$ matrix in the $\ket{m}_q$ eigenbasis $\ket{\I{\pm}\frac{1}{2}}$ 
and can be cast as a linear combination of the unit and Pauli matrices:
\begin{equation}
\hat{\varrho}_{q}(t)\equiv
\left(\begin{smallmatrix}
\matr{+\frac{1}{2}}{\hat{\varrho}_{q}(t)}{+\frac{1}{2}} & \matr{+\frac{1}{2}}{\hat{\varrho}_{q}(t)}{-\frac{1}{2}}\\
\matr{-\frac{1}{2}}{\hat{\varrho}_{q}(t)}{+\frac{1}{2}} & \matr{-\frac{1}{2}}{\hat{\varrho}_{q}(t)}{-\frac{1}{2}}\end{smallmatrix}\right)
=\frac{1}{2}\left[\hat{1}+\vecb{\wp}(t)\cdot\hat{\vecb{\sigma}}\right].
\label{denmat1} 
\end{equation}
The Bloch vector $\vecb{\wp}(t)=2\vecb{J}(t)$ represents an instantaneous quantum expectation value of the normalized spin polarization, so
\begin{equation}
\hat{\varrho}_{q}(t)=
\left(\begin{array}{cc}
\frac{1}{2}+J_z(t) & J_x(t)-\ui J_y(t)\\
J_x(t)+\ui J_y(t) & \frac{1}{2}-J_z(t)
\end{array}\right).
\label{demat2}
\end{equation}
For the qubit survival probability we immediately obtain
\begin{equation}
P_q(t)=\frac{1}{2}-J_z(t).
\label{survq1}
\end{equation}

Another measure that will be applied is the degree of purity of the ${N=1}$ state \eqref{demat2}.
It is characterized by 
\begin{equation}
{\rm Tr}\!\left[\hat{\varrho}_{q}(t)^2\right]\I{=}\frac{1}{2}\I{+}2\left[J_x(t)^2\!\I{+}J_y(t)^2\!\I{+}J_z(t)^2\right]\I{=}\frac{\wp(t)^{2}\I{+}1}{2},
\label{puri}
\end{equation}
which always takes the same value (lying within the interval $[\frac{1}{2},1]$) as ${\rm Tr}[\hat{\varrho}_{b}(t)^2]$. 
The normalized purity ${\wp(t)=\abs{\vecb{\wp}(t)}}$ is connected with the polarization of the qubit system and changes within the interval ${\wp\in[0,1]}$. 
The limits ${\wp=0}$ and 1 correspond to maximally mixed (completely unpolarized) and pure (maximally polarized) states, respectively.
This quantity will be used as a measure of the decoherence induced on both qubit and field subsystems by their mutual interaction during the evolution.
It also quantifies the fulfilment of the approximate adiabatic separation explained in Sec.\,\ref{Clasgen}.
Perfect adiabatic separation implies ${\wp=1}$, while decreasing values ${\wp<1}$ indicate increasing violation of this approximation.

Let us finally mention some specific features of quantum dynamics with the initial state from Eq.\,\eqref{evo}. 
The mean of the energy distribution \eqref{surv} in this case is independent of the choice of parameters ${(\lambda,\delta,\mu,\gamma)}$ of the final Hamiltonian.
We always obtain ${\ave{E}=-\omega Rj}$, hence ${\varepsilon=-\frac{1}{2}}$, which for ${\gamma=1}$ is precisely the energy associated with the stationary point at $(q,p)\I{=}(0,0)$.
We know that for ${\lambda>\lambda_{\rm c}}$ or~$\lambda'_{\rm c}$ this is always an ESQPT critical energy, the type of the spectral singularity being linked the type of the respective superradiant phase, see Figs.\,\ref{stacpoints1}, \ref{stacpoints2}(a) and \ref{stacpoints2}(b).
For the quenches with ${\mu=0}$ and ${\delta=+1}$, the initial state $\ket{m\I{=}-j}_{q}\I{\otimes}\ket{n\I{=}0}_{b}$ coincides with a permanent eigenstate  of the Jaynes-Cummings Hamiltonian, so there is no quench dynamics as the system remains frozen.
This can be applied in a two-step realization of quantum quench protocols to general parity-conserving Hamiltonians with ${\lambda>0}$ and ${\delta\neq 1}$: 
In the first step, the system is set to ${\delta=+1}$ and pushed to the desired value of $\lambda$, and in the second step $\delta$ jumps to its requested value while $\lambda$ is kept constant.

\subsection{Parity conserving case}
\label{Dyncons}

\begin{figure}[t]
\includegraphics[width=\linewidth]{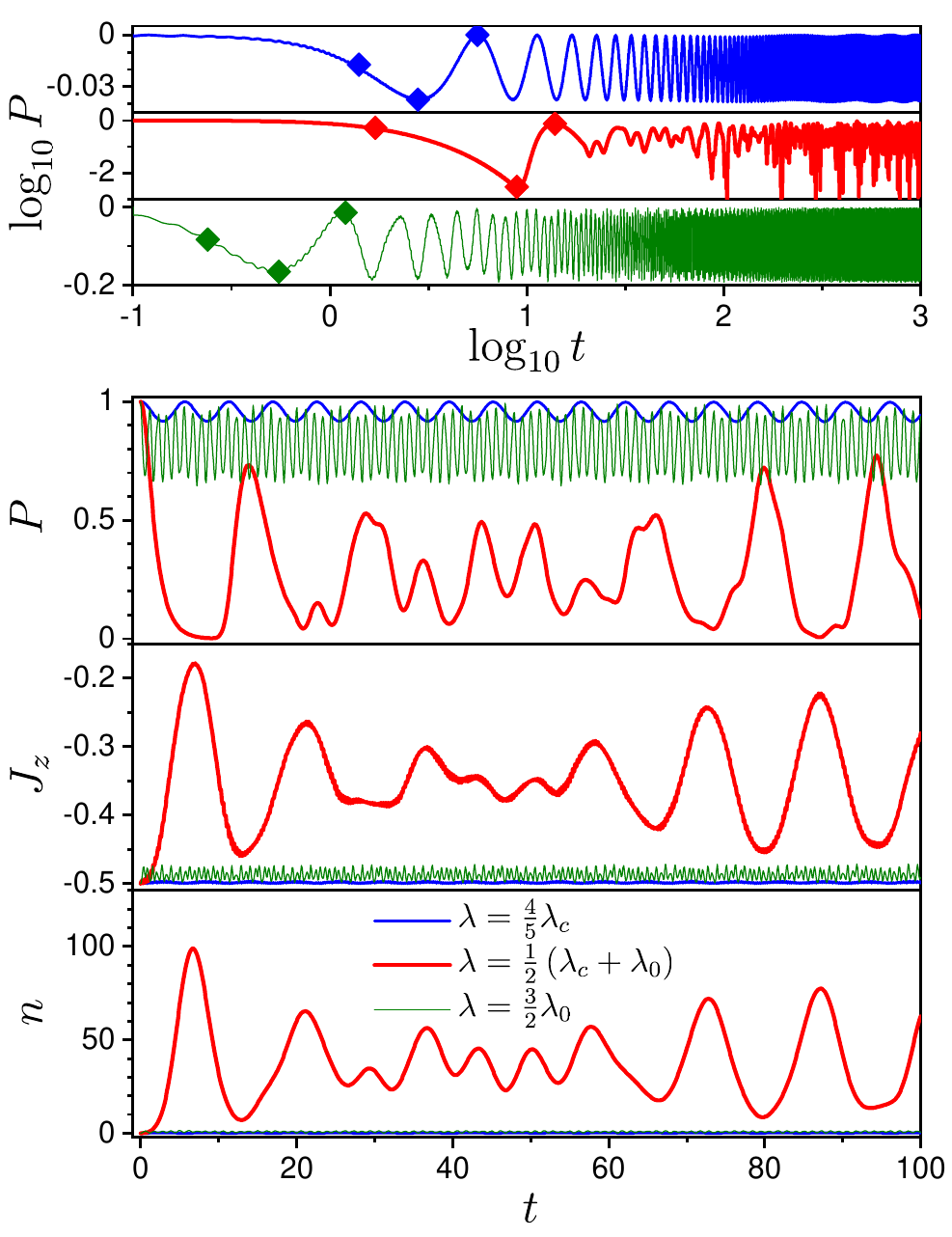}
\caption{Time dependencies $P(t)$, $J_z(t)$ and $n(t)$ in the evolution \eqref{evo} by the parity-conserving (${\mu=0}$) Hamiltonian with ${\lambda=0.8\lambda_{\rm c}}$ (the normal phase), ${\lambda=(\lambda_{\rm c}\I{+}\lambda_{0})/2}$ (the first superradiant phase), and ${\lambda=1.5\lambda_0}$ (the second superradiant phase).
The other parameters are ${\delta=0.5}$, ${R=100}$ and ${\omega=1=N}$. 
The survival probability is shown in the linear scale (in the main plot) as well as in the log-log scale for a longer time interval (in the upper plots).
Significant departures of all quantities from their ${t=0}$ values are observed only in the first superradiant phase.
The diamonds in the log-log plots correspond to the time instants visualized in Fig.\,\ref{wigner}.}
\label{timedep}
\end{figure}

\begin{figure}[t]
\includegraphics[width=\linewidth]{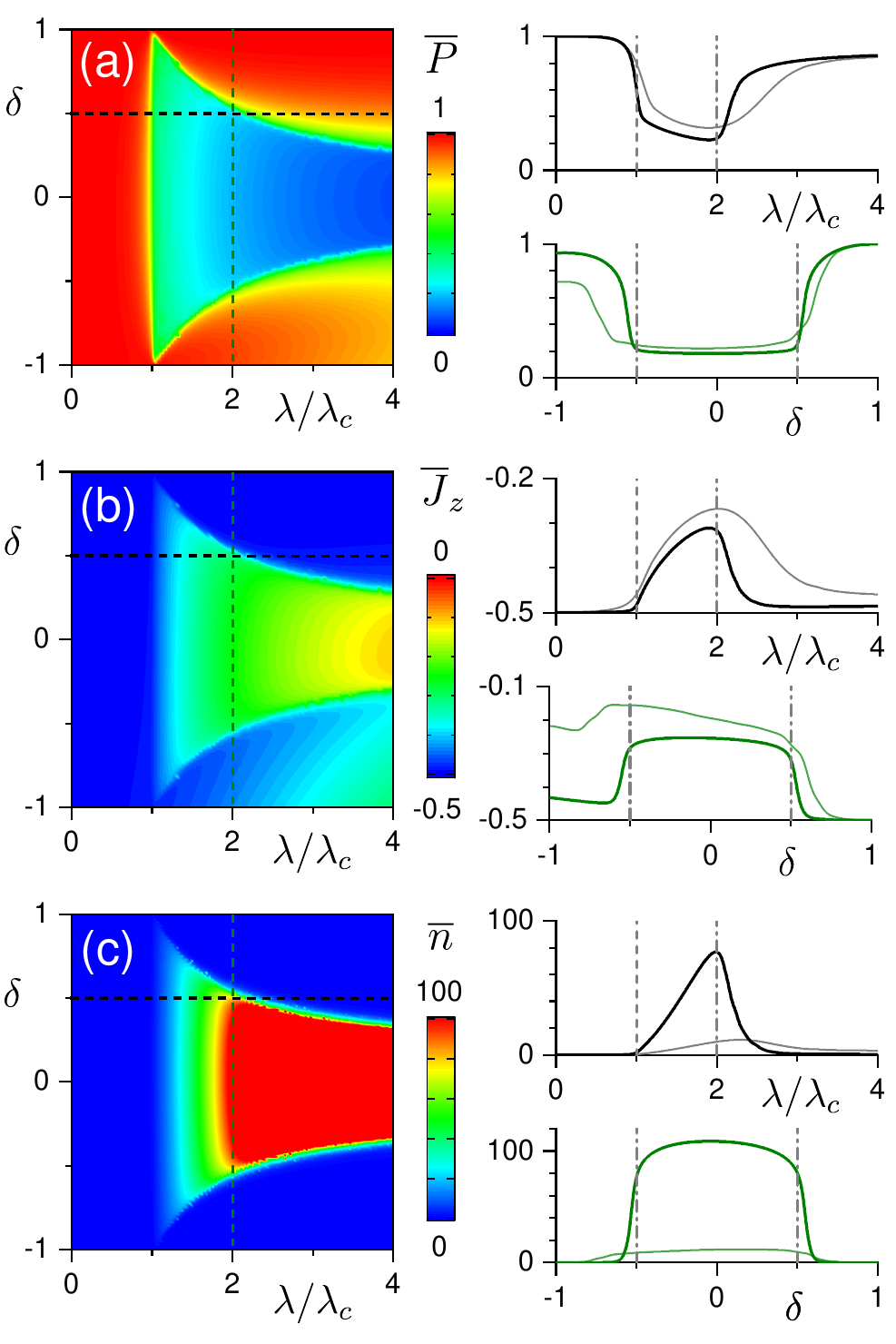}
\caption{Infinite-time averages  in the evolution \eqref{evo} by the parity-conserving (${\mu=0}$) Hamiltonian with ${N=1}$.
Panels show the averages \eqref{avera} for (a) the survival probability $P(t)$, (b) the expectation value of the quasispin $z$-projection $J_z(t)$, and (c) the expectation value of the boson number $n(t)$.
The planar plot in each panel depicts the averages in the $\lambda\times\delta$ plane for ${R=100}$ (cf.\,Fig.\,\ref{phase}), the graphs show dependencies along the two cuts (dashed lines in the planar plot) for ${R=10}$ (thin curves) and ${R=100}$ (thick curves).
All dependencies are locally smoothed in parameters $\lambda$ and $\delta$ to get rid of rapid oscillations.
Vertical dashed and dash-dotted lines in the graphs mark the values ${\lambda=\lambda_{\rm c}}$ and $\lambda_0$ from Eq.\,\eqref{lac}.
}
\label{timave}
\end{figure}

\begin{figure}[t]
\includegraphics[width=\linewidth]{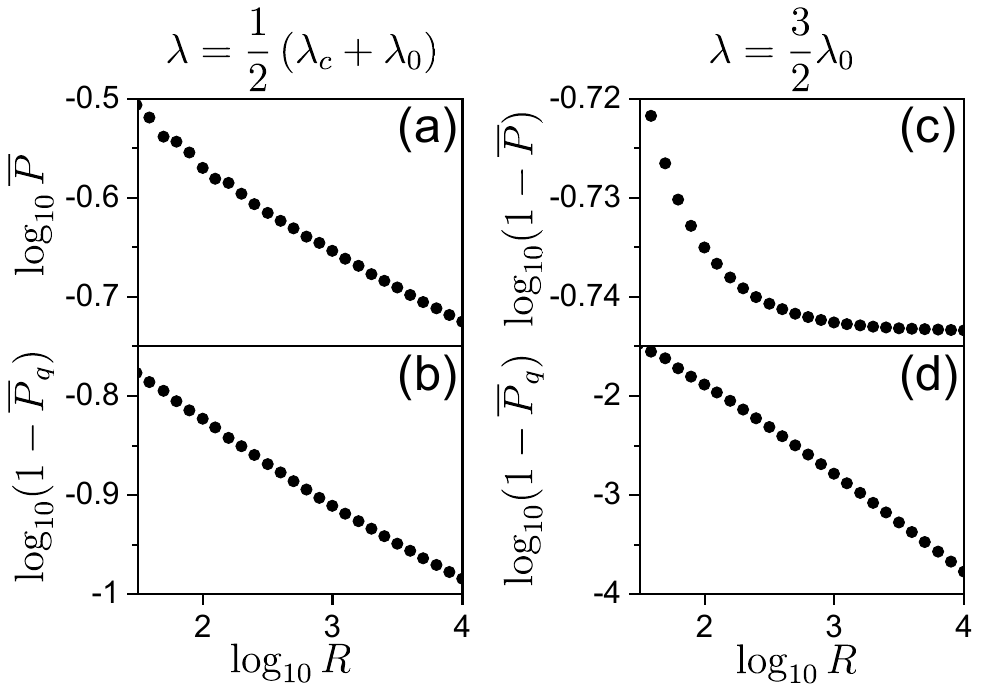}
\caption{Variations of averaged survival probabilities $\overline{P}$ (the qubit-field initial state) and $\overline{P_q}$ (the qubit initial state) with increasing size parameter $R$.
The parameters of the parity-conserving Hamiltonian are the same as in Fig.\,\ref{timedep}.
Panels (a),(b) and (c),(d), respectively, correspond to the selected values of $\lambda$ in the first and second superradiant phases.
The ${R\to\infty}$ behavior in these phases differs: while in the ${\rm S}_2$ phase both survival probabilities increase towards unity, in the ${\rm S}_1$ phase $\overline{P}$ decreases to zero but $\overline{P_q}$ increases to unity.
Note that the decreasing dependencies in panels (a),(b) and (d) are roughly algebraic, but the one in panel (c) is much slower.}
\label{loglog}
\end{figure}

\begin{figure}[t]
\includegraphics[width=\linewidth]{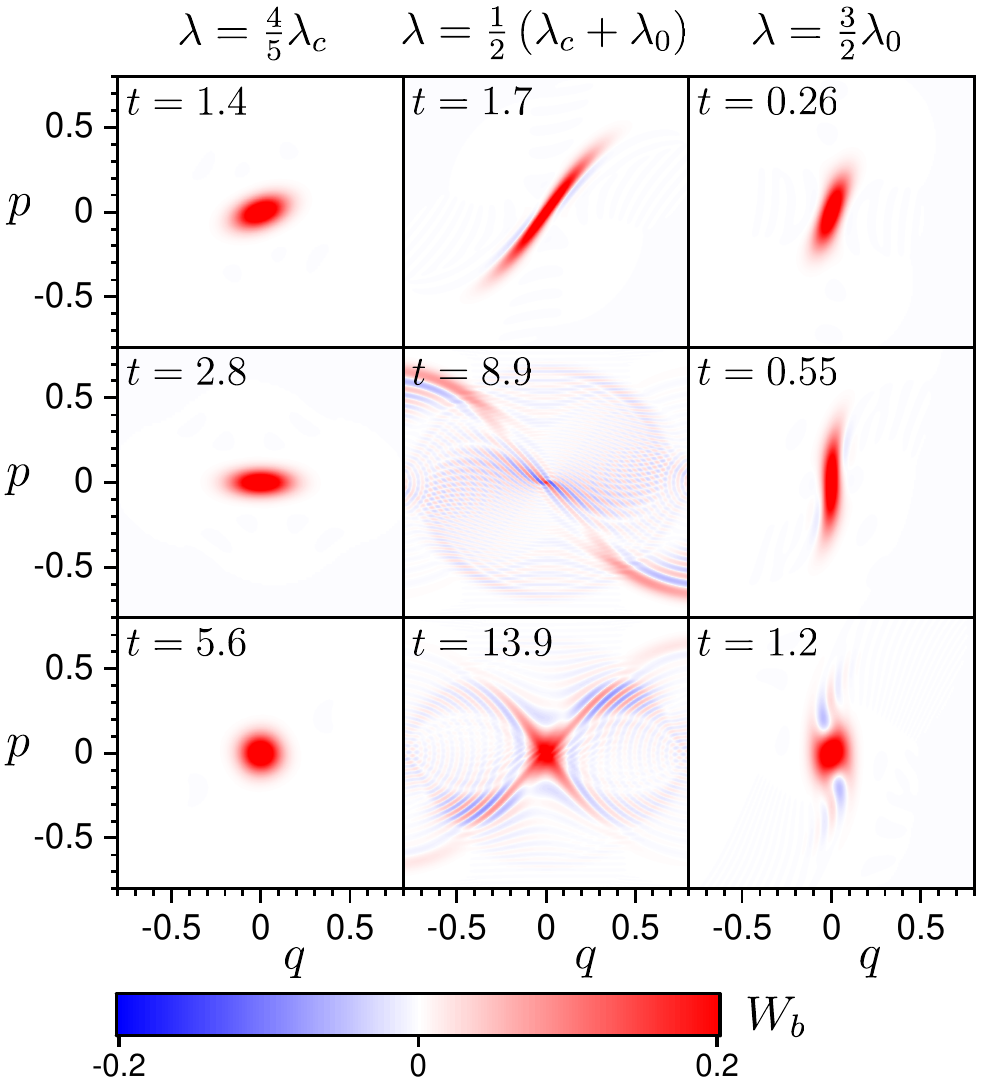}
\caption{Snapshots of the exact Wigner quasiprobability distribution in the phase space following evolution \eqref{evo} by the parity-conserving Hamiltonians with the same parameters as in Fig.\,\ref{timedep}.
The three columns correspond to the values of $\lambda$ in the normal, first superradiant, and second superradiant phases.
The rows correspond to three time instants, which were chosen in the very initial stage of the evolution, in the first dip of the survival probability, and the in first approximate revival of the initial state; see the diamonds in the upper plots of Fig.\,\ref{timedep}.
We clearly observe the stabilizing character of the $(q,p)\I{=}(0,0)$ stationary point in the normal and second superradiant phases, cf.\,Fig.\,\ref{stacpoints1}.
The Wigner distributions have been calculated using the QuantumOptics.jl framework \cite{Kra18}.}
\label{wigner}
\end{figure}

We first consider quantum dynamics induced by the parity-conserving Rabi Hamiltonians with ${\mu=0}$.
Time dependencies of the survival probability and some expectation values are shown in Fig.\,\ref{timedep}.
The choice of an initial state with a fixed (positive) value of parity $\hat{\Pi}$ from Eq.\,\eqref{parit} and its conservation during the evolution implies vanishing of some expectation values, namely 
\begin{equation}
J_x(t)=J_y(t)=q(t)=p(t)=0.
\label{zero}
\end{equation}
Indeed, for any state $\ket{\Psi(t)}$ satisfying $\hat{\Pi}\ket{\Psi(t)}\I{=}\pm\ket{\Psi(t)}$ we obtain ${\matr{\Psi(t)}{\hat{X}}{\Psi(t)}\I{=}-\matr{\Psi(t)}{\hat{X}}{\Psi(t)}}$ for ${\hat{X}\I{=}\hat{J}_x}$, $\hat{J}_y$, $\hat{q}$ and $\hat{p}$.
Nonzero expectation values of these quantities can be observed only for states with mixed parity.

The case of ${N=1}$ in connection with the parity conservation leads to additional special features.
In particular, Eqs.\,\eqref{puri} and \eqref{zero} imply that ${\wp(t)=2\abs{J_z(t)}}$.
So the projection $J_z(t)$ fully determines not only the qubit survival probability, Eq.\,\eqref{survq1}, but also the purity. 
Moreover, since the coefficients $c_{mn}(t)$ in Eq.\,\eqref{evo} are nonzero only for ${m=-\frac{1}{2}}$ combined with ${n=0,2,4,\dots}$ and ${m=+\frac{1}{2}}$ combined with ${n=1,3,5,\dots}$, we obtain the following relation between the survival probabilities: 
\begin{equation}
P(t)=P_b(t)\leq P_q(t).
\label{eqineq}
\end{equation}
We see that the infinite-time averages ${\overline{P}=\overline{P}_b}$ set a lower bound for $\overline{P}_q$.
For ${R\to\infty}$, when the factorization ${P(t)=P_q(t)P_b(t)}$ takes place, there are two possibilities:
Either the ${R\to\infty}$ limit of ${\overline{P}=\overline{P}_b}$ is nonzero and the limit of $\overline{P}_q$ is unity, or the limit of ${\overline{P}=\overline{P}_b}$ is zero and the limit of $\overline{P}_q$ is arbitrary.
As demonstrated below, the case ${\overline{P}_q\to 1}$ applies to our ${N=1}$ initial state $\ket{\Psi(0)}={\ket{m\I{=}-\frac{1}{2}}_q\otimes\ket{n\I{=}0}_b}$ involving the field vacuum.
In contrast, it can be shown that the case ${\overline{P}\to 0}$ applies, for instance, to non-vacuum initial states ${\ket{m\I{=}-\frac{1}{2}}_q\otimes\ket{n\I{>}0}_b}$ with scaled energy ${\varepsilon>-\frac{1}{2}}$.

The time dependencies shown in Fig.\,\ref{timedep} indicate that the Hamiltonians in the normal and second superradiant phases (see Fig.\,\ref{phase}) induce only small deviations of the quantities observed from their initial values.
This is true mostly for the average boson number $n(t)$ and the average quasispin $z$-projection $J_z(t)$, hence also for the qubit survival probability $P_q(t)$ and purity $\wp(t)$, but it applies to a large extent also to the overall survival probability $P(t)$. 
Significant deviations from the initial values are seen only for the evolution by the Hamiltonian in the first superradiant phase.
The log-log plots of the survival probability expand the short-time evolution and visualize a longer time interval in which the regime with saturated fluctuation measures is reached.

Figure \ref{timave} depicts the infinite-time averages \eqref{avera} of the quantities from Fig.\,\ref{timedep}.
These are plotted for ${R=100}$ in the parameter plane $\lambda\times\delta$ of the parity-conserving Hamiltonian, and along its two cuts for ${R=10}$ and 100.
The dependencies shown in these plots are smoothed (by the moving average method) to get rid of rapid oscillations with control parameters that reflect specific $R$-dependent patterns of avoided level crossings in the corresponding spectra.
It turns out that for large~$R$ the smoothed infinite-time averages in Fig.\,\ref{timave} enable us to distinguish the three quantum phases from Sec.\,\ref{Clascons}.
The observations can be summarized as follows:

(a) In the normal phase, i.e., for ${\lambda\in[0,\lambda_{\rm c}]}$ (domain N in Fig.\,\ref{phase}), the averaged overall survival probability is ${\overline{P}\approx 1}$, while the averaged expectation values of the number of bosons and the quasispin $z$-projection are given by ${\overline{n}\approx 0}$ and ${\overline{J_z}\approx-\frac{1}{2}}$. 
So the averaged qubit survival probability and purity ${\overline{P_q}\approx\overline{\wp}\approx 1}$ (not shown in the figure).
In this parameter domain, the noninteracting ground state $\ket{m\I{=}-j}_{q}\I{\otimes}\ket{n\I{=}0}_{b}$, which is the initial state of the quantum evolution \eqref{evo}, coincides to a high degree of accuracy with the ground state of the actual Hamiltonian $\hat{H}$.
So the system is in a nearly stationary state.
In the infinite-size limit ${R\to\infty}$, when the ground-state criticality becomes sharp, these statements would be valid in the absolute sense. 

(b) In the first superradiant phase, for ${\lambda\in(\lambda_{\rm c},\lambda_0(\delta)]}$ (domain ${\rm S}_1$ in Fig.\,\ref{phase}), the average survival probability $\overline{P}$ is considerably reduced. 
At the same time, the averaged number of bosons $\overline{n}$ grows with $\lambda$ (almost independently of $\delta$) to values ${\approx R}$ and the averaged quasispin projection $\overline{J_z}$ deviates from the spin-down value~$-\frac{1}{2}$. 
The averaged qubit survival probability and purity in finite-size cases decrease correspondingly below 1, nevertheless they remain well above 0.
For increasing $R$ we observe that both $\overline{P_q}$ and $\overline{\wp}$ increase towards 1 while the averaged value of the overall survival probability $\overline{P}$ decreases down to 0.
This means that with increasing $R$ the qubit in the ${\rm S}_1$ phase more and more preserves its initial state, but the field characterized by growing $\overline{n}$ gradually deviates from its vacuum state. 
This is demonstrated in Fig.\,\ref{loglog}, whose panels~(a) and~(b) show dependencies of the overall and qubit survival probabilities on the size parameter (up to $R=10^4$) for a single point in the middle of the ${\rm S}_1$ phase.
We observe approximate algebraic dependencies ${\overline{P}\propto N^{-\alpha_1}}$ and ${1-\overline{P_q}\propto N^{-\beta_{1}}}$, with exponents ${\alpha_1>0}$ and ${\beta_1>0}$ given (for the selected~$\lambda$) by the slopes of the respective plots.
In the limit ${R\to\infty}$ we get ${\overline{P}\to 0}$ and ${\overline{P_q}\to 1}$.

(c) In the second superradiant phase, that is for ${\lambda\in(\lambda_{0}(\delta),\infty)}$ (domains ${\rm S}_2$ and ${\rm S}'_2$ in Fig.\,\ref{phase}) the situation becomes closer to the normal phase.
The average survival probability $\overline{P}$ takes values closer to 1 again, the averaged number of bosons $\overline{n}$ returns to ${\approx 0}$, and the averaged quasispin projection $\overline{J_z}$ gets closer to the initial value $-\frac{1}{2}$ (increasing the averaged qubit survival probability and purity back to values near 1). 
The effect is stronger and sharper for larger $R$.
Hence, despite the strong qubit-field interaction in both ${\rm S}_2$ and ${\rm S}'_2$ phases, the whole initial product state is preserved to a large extent even during an asymptotically long evolution.
The increase of stabilization with $R$ is illustrated, for a single point of the ${\rm S}_2$ phase, in panels (c) and (d) of Fig.\,\ref{loglog}.
For the qubit survival probability we observe an approximately algebraic dependence ${1-\overline{P_q}\propto N^{-\beta_{2}}}$ with an exponent ${\beta_2>0}$, implying therefore  the ${R\to\infty}$ behavior ${\overline{P_q}\to 1}$.
The overall survival probability $\overline{P}$ increases in a nonalgebraic way, leaving the ${R\to\infty}$ behavior indeterminable.

The explanation of these phenomena follows directly from the stability properties of the ${(q,p)=(0,0)}$ stationary point discussed at the end of Sec.\,\ref{Clascons}.
In order to see this, we recall the results of Ref.\,\cite{Klo21}, where the quantum quench dynamics was efficiently approximated by the classical evolution of the Wigner quasiprobability distribution in the phase space.
In the present system, the Wigner distribution is defined by
\begin{eqnarray}
W_b(q,p,t)=\frac{1}{\pi}\int_{-\infty}^{+\infty}\!\!\!\!
dx'\,\matr{x\I{+}x'}{\hat{\varrho}_b(t)}{x\I{-}x'}\, e^{-2\ui px'}
\qquad\label{Wig}\\
=\frac{1}{\pi}\sum_{n,n'}\int_{-\infty}^{+\infty}\!\!\!\!
dx'\,\scal{x\I{+}x'}{n}\matr{n}{\hat{\varrho}_b(t)}{n'}\scal{n'}{x\I{-}x'}\, e^{-2\ui px'},
\nonumber
\end{eqnarray}
where the matrix elements $\matr{n}{\hat{\varrho}_b(t)}{n'}$ can be substituted from Eq.\,\eqref{rhob} and scalar products of the type $\scal{x}{n}$ coincide with eigenfuctions of the one-dimensional harmonic oscillator.
The survival probability of the initial field state at any time reads as
\begin{equation}
P_b(t)=2\pi\iint_{-\infty}^{+\infty}dq\,dp\,W_b(q,p,t)W_b(q,p,0).
\label{Wover}
\end{equation}
This formula is exact if the Wigner distribution at time~$t$ is calculated, as assumed in Eq.\,\eqref{Wig}, from the correct field state $\hat{\varrho}_b(t)$.
A semiclassical approximation of $W_b(q,p,t)$ and the survival probability \eqref{Wover} can be obtained if the initial distribution $W_b(q,p,0)$ associated with the vacuum state $\ket{n=0}_b$ is evolved by applying only the classical equations of motions.
It is clear that this approximation can yield reasonable results only for moderate times.
Nevertheless, the connection to classical dynamics invoked in these considerations gives us intuitive insight into the roots of qualitatively different dynamical responses observed in quantum quenches to various parameter domains.

The exact evolution of the the Wigner distribution in various vacuum phases of the parity-conserving Hamiltonian is illustrated in Fig.\,\ref{wigner}.
We see that the initial distribution $W_b(q,p,0)$ is always centered right at the ${(q,p)=(0,0)}$ stationary point. 
Since a~certain fraction of classical orbits near the stationary point is slow in leaving the initial phase-space domain, the decay of the initial state must be partly suppressed.
The degree of this suppression however strongly depends on the stability properties of the $(0,0)$ point in various parameter domains:
In the normal phase, the $(0,0)$ point (the global minimum of the Hamiltonian) is very stable and therefore the evolution of the Wigner distribution is marginal.
In the first superradiant phase, the $(0,0)$ point (a saddle point of the Hamiltonian) is unstable, so the dissipation of the lateral parts of $W_b(q,p,t)$ characterizes the departure of the field from the vacuum state and reduces the overlap with $W_b(q,p,0)$.
Nevertheless, the central part of the distribution, which is little affected for moderate times, still makes the field survival probability larger than in cases when no stationary point is present.
Finally, in the second superradiant phase, the $(0,0)$ point (a local maximum of the Hamiltonian function) is stable again, hence the evolving Wigner distribution remains close to its initial form.

We stress that for ${N=1}$ and in the parity-conserving case, according to Eq.\,\eqref{eqineq}, the field survival probability $P_q(t)$ in Eq.\,\eqref{Wover} coincides with the overall survival probability $P(t)$ and sets a lower bound for the qubit survival probability $P_q(t)$. 
The calculations presented in Fig.\,\ref{loglog} show that the convergence of $P_q(t)$ to unity is very fast in the normal and second superradiant phases, which can be connected with the stability of the vacuum state.
On the other hand, in the first superradiant phase the increase of $P_q(t)$ is much slower as $P_b(t)$ sets a less stringent lower bound.

\subsection{Parity violating cases}
\label{Dynviol}

\begin{figure}[t]
\includegraphics[width=\linewidth]{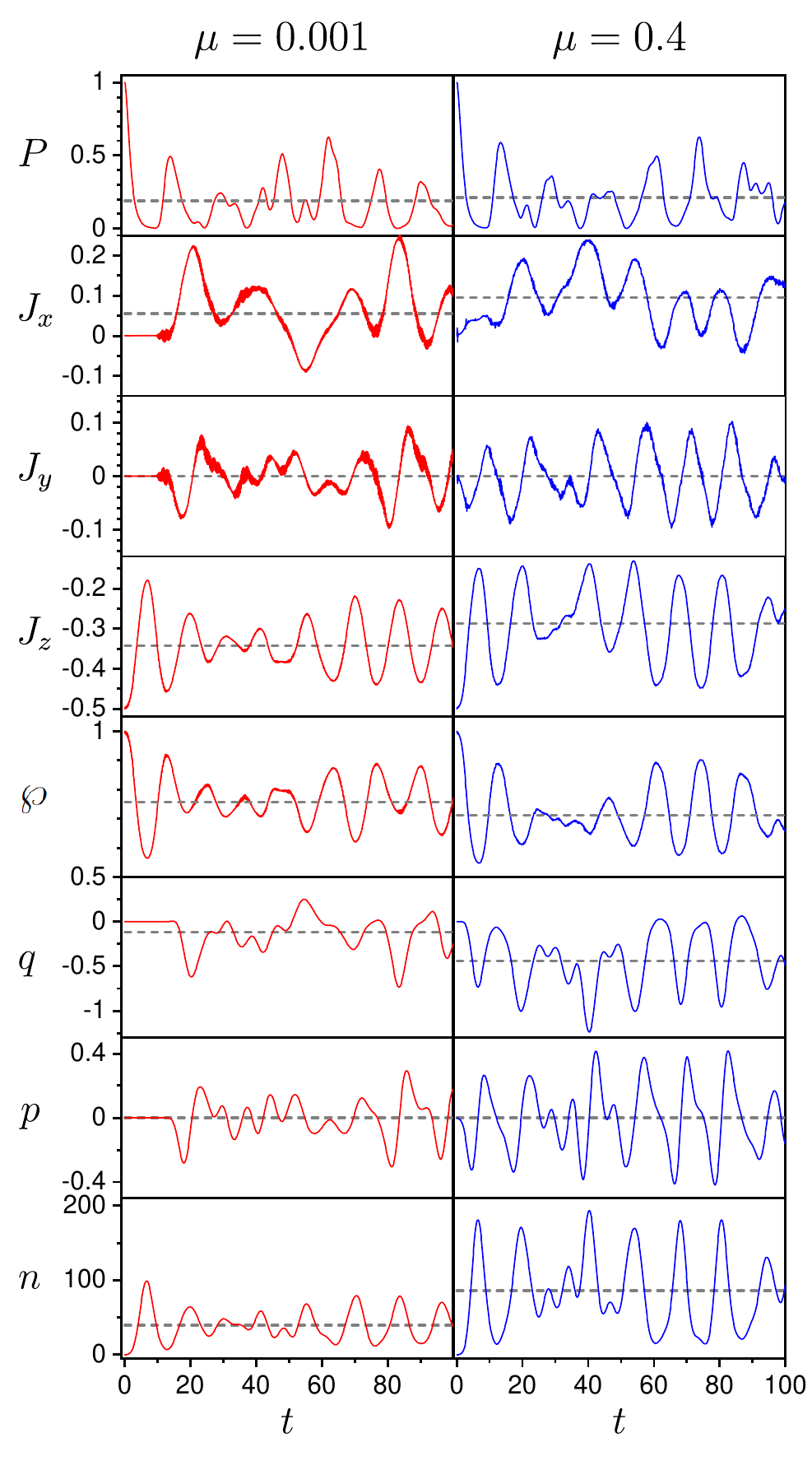}
\caption{Various time dependencies corresponding to the evolution \eqref{evo} by the parity-violating Hamiltonian with two values of $\mu$ (left and right columns). 
The other parameters are: ${\gamma=1}$, ${\lambda=0.75}\I{=}\frac{1}{2}(\lambda_{\rm c}\I{+}\lambda_0)$, ${\delta=0.5}$, ${R=100}$, and ${\omega=1=N}$.
The rows from top to bottom depict the overall survival probability, quantities assigned to the qubit (expectation values of the three quasispin projections and the purity) and to the field (expectation values of the coordinate, momentum and boson number).
The dashed horizontal lines represent infinite-time averages.}
\label{timedep2}
\end{figure}

\begin{figure}[t]
\includegraphics[width=\linewidth]{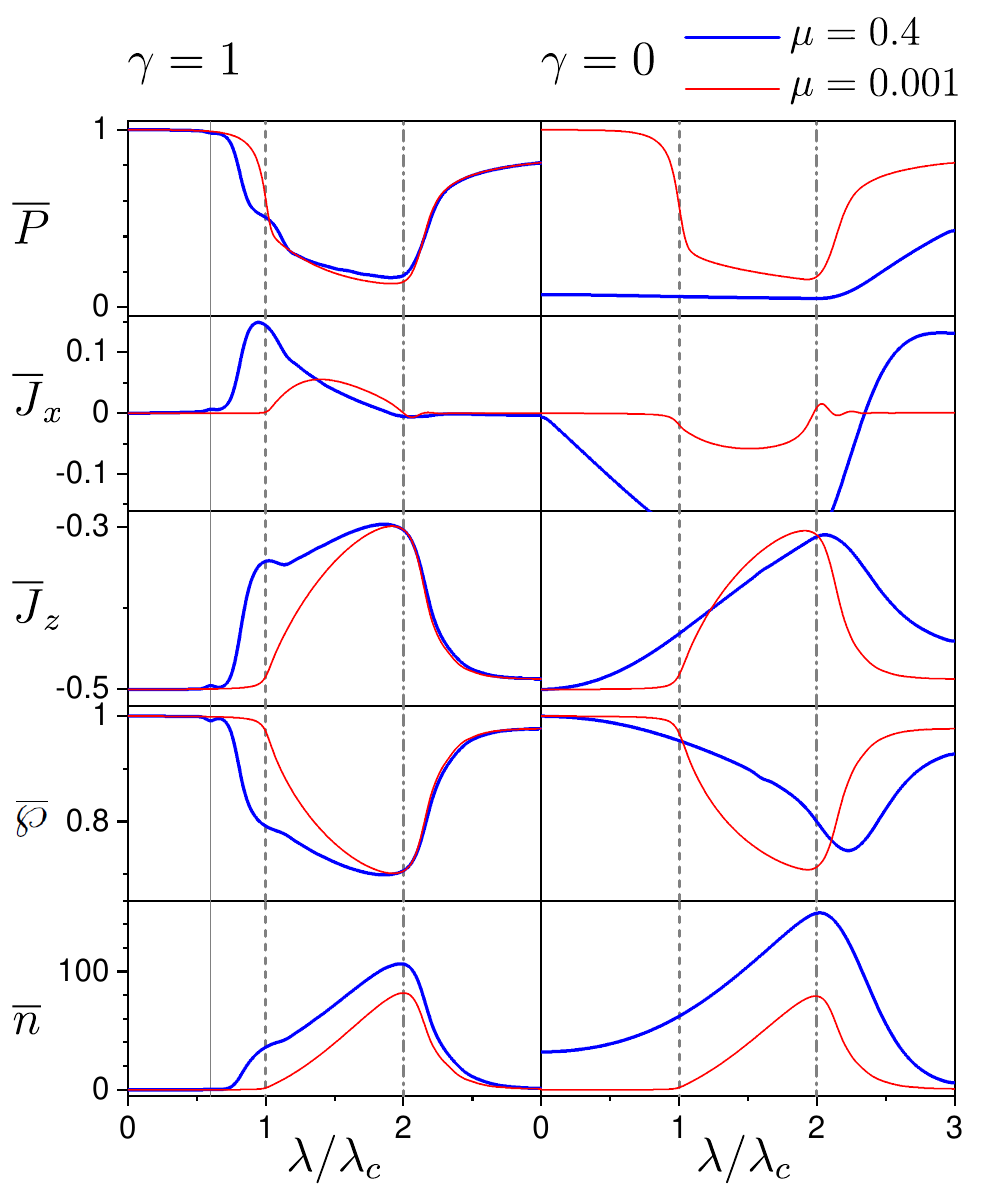}
\caption{
Infinite-time averages in the evolution \eqref{evo} by the parity-violating  Hamiltonians with the indicated values of $\mu$ and $\gamma$ for $\lambda\in[0,3\lambda_{\rm c}]$, ${\delta=0.5}$, ${N=1}$, and ${R=100}$.
The quantities depicted are the same as in Fig.\,\ref{timedep2} except $J_y$, $q$ and $p$. 
The left and right columns correspond to ${\gamma=1}$ and~0, respectively, the two values of $\mu$ are distinguished by line types (cf.\,Fig.\,\ref{timedep2}).
All dependencies are locally smoothed in parameter $\lambda$.
Vertical dashed and dash-dotted lines mark the values ${\lambda=\lambda_{\rm c}}$ and~$\lambda_0$ from Eq.\,\eqref{lac}.
The full vertical line corresponds to the first-order QPT point $\lambda_{\rm c}'$ for ${\mu=0.4}$ and ${\gamma=1}$.
}
\label{timave2}
\end{figure}

For parity violating Hamiltonians with ${\mu\neq 0}$, the condition \eqref{zero} is generally broken. 
For instantaneous expectation values we expect ${J_x(t),J_y(t),q(t),p(t)\neq 0}$.
Also the infinite-time averages of the $x$-polarization and coordinate in general satisfy ${\overline{J_x},\overline{q}\neq 0}$. 
However, the infinite-time average of the $y$-polarization and momentum always yields 
\begin{equation}
\overline{J}_y=\overline{p}=0,
\label{null} 
\end{equation}
which is a consequence of Eq.\,\eqref{aaver} with the substitution ${\matr{E_i}{\hat{J}_y}{E_i}=\matr{E_i}{\hat{p}}{E_i}=0}$ following from the reality of the Hamiltonian matrix elements in the basis $\ket{m,n}$ and the subsequent reality of coefficients $c_{imn}$ in the expansion $\ket{E_i}\I{=}\sum_{m,n}c_{imn}\ket{m,\!n}$.

Figure \ref{timedep2} exemplifies time dependencies of several quantities for two values of the parity-violating interaction parameter, a small one, ${\mu=10^{-3}}$, and a medium one, ${\mu=0.4}$ (strengths of interactions are in general compared with parameter $\omega$ weighting the noninteracting part of the Hamiltonian).
In this figure we consider only the case ${\gamma=1}$ with the first-order ground-state QPT, see Fig.\,\ref{stacpoints2}(a).
The parity-conserving interaction parameter~$\lambda$ is taken in the middle of the first superradiant phase, right on the halfway between~$\lambda_{\rm c}$ and $\lambda_0$.
Otherwise we fix ${\delta=0.5}$ and ${R=100}$.
The figure displays the expectation values of all quasispin components, the boson number, its quadratures, and also the overall survival probability and purity.
The infinite-time averages of the respective quantities are drawn by horizontal lines, so one immediately verifies the fulfillment of Eq.\,\eqref{null}.
The amplitudes of temporal variations of the quasispin components do not significantly depend on the size of $\mu$, so even a very weak parity-violating perturbation produces a relatively large effect in nonzero values of $J_x(t)$ and $J_y(t)$.

In the left column of Fig.\,\ref{timedep2} we note a sharp onset of $J_x(t)$ and $J_y(t)$ at a certain time instant, soon followed by a similar onset of $q(t)$ and $p(t)$.
This sudden violation of parity in the evolving states of the qubit and the field takes place only for small values of the parameter~$\mu$.
We note that in a finite system the expectation values of the respective quantity before its onset are not exactly zero but take fluctuating values much smaller than those after the onset.
This surprising phenomenon is not important in the present context, but may have rather interesting consequences for short-time dynamics.
The effect is due to an asymmetric interference of two wave packets that arise from the initial state (a single Gaussian centered at ${q=0}$) and move across the two nearly symmetric (for small $\mu$) Hamiltonian minima at ${q<0}$ and ${q>0}$.
We are currently working on a more detailed analysis, which will be reported elsewhere.

The infinite-time averages \eqref{avera} are presented in Fig.\,\ref{timave2} along the ${\delta=0.5}$ cut of the $\lambda\times\delta$ plane.
We show the same quantities as in Fig.\,\ref{timedep2}, except $J_y$ and $p$, which identically satisfy Eq.\,\eqref{null}, and $q$.
We again have ${R=100}$ and the same pair of values of the parameter $\mu$.
Besides the results for ${\gamma=1}$ (parity violation via a qubit-field interaction and an external drive of the field) we present also those for ${\gamma=0}$ (parity violation without an external drive), when the system has no ground-state QPT and the vacuum state $(q,p)\I{=}(0,0)$ is not a stationary point for any value of parameter~$\lambda$.
In analogy to the parity-conserving case summarized in Fig.\,\ref{timave}, the infinite-time averages in the parity-violating case exhibit large and rapid oscillations with $\lambda$, so the plots in Fig.\,\ref{timave2} display appropriately smoothed dependencies.

The infinite-time averages demonstrate that even the presence of parity-violating terms in the Hamiltonian does not destroy the initial-state stabilization effect described above.
This effect remains present particularly for large values of $\lambda$, i.e., in a deep superradiant regime.
Specific observations deduced from Fig.\,\ref{timave2} can be summarized as follows:

(i) For the small value of the parity-violating interaction strength $\mu$, independently of $\gamma$, the averaged survival probability, purity as well as the expectation values of the boson number and the quasispin $z$-projection show very similar behavior as in the parity-conserving case with ${\mu=0}$ (cf.\,Fig.\,\ref{timave}).
A qualitatively new feature in Fig.\,\ref{timave2} is a small nonzero value of the averaged quasispin $x$-projection.
We observe that its sign depends on the value of $\gamma$.

(ii) For the larger value of $\mu$ and ${\gamma=1}$, the averages distinguish not three, but rather four vacuum phases described in Sec.\,\ref{Clasviol}.
These are the normal phase below the first-order QPT at ${\lambda=\lambda_{\rm c}'}$, the zeroth superradiant phase between $\lambda_{\rm c}'$ and $\lambda_{\rm c}$, and the first and second superradiant phases above $\lambda_{\rm c}$ and $\lambda_0$, respectively, see Fig.\,\ref{stacpoints2}(a).
The distinction would become sharper for even larger values of $R$.
The stabilization of the initial state is again strong in the normal and second superradiant phases.
A partly increased overall survival probability \eqref{surv} is observed also in the zeroth superradiant phase, where the local stability properties of the vacuum are the same as in the normal phase, but in the zeroth superradiant phase the stabilization of the qubit itself, expressed by the averaged probability \eqref{survq1}, is weaker, comparable to the first superradiant phase.

(iii) For the larger value of $\mu$ and ${\gamma=0}$, the infinite-time averages show qualitatively different dependencies on~$\lambda$.
In this case, the initial-state stabilization is observed mostly for large values of $\lambda$, for which the relative role of the parity-violating term in the Hamiltonian is reduced and the system gets close to the second superradiant phase of the parity-conserving model.
This view is supported by the convergence of the ESQPT critical energies in Fig.\,\ref{stacpoints2}(c) to those of Fig.\,\ref{stacpoints2}(b) as $\lambda$ increases.

We therefore conclude that an efficient stabilization effect occurs even for parity-violating Hamiltonians, independently of the form and strength of the parity-violating interactions.
The phenomenon takes place if the parity-conserving interactions are dominant and the degree of stabilization increases with $\lambda$.

\section{Conclusions}
\label{Clouseau}

In this work, we have performed a QPT and ESQPT analysis of an extended Rabi model including both parity-conserving and parity-violating interactions.
The model describes a coupled qubit-field system in which sharp quantum critical effects occur in spite of a very small size of the qubit subsystem, even in the ${N\I{=}1}$ case.
It has been clarified that a necessary condition for criticality in this model is infinite imbalance between both system components (asymptotically large ratio $R$ between the qubit and field elementary excitation energies), so that the infinite-size limit is realized solely via the field subsystem whose bosonic quanta in typical eigenstates yield average numbers ${\ave{\hat{n}}\to\infty}$.

Using this assumption, we have disclosed rather complex phase structures associated with both parity-conserving and parity violating versions of the model.
Some parameter subsets contain either the first-order, or second-order ground-state QPTs of the superradiant type, some others show only continuous evolutions of the ground state. 
Numerous ESQPTs appear in the excited domain and generate various types of nonanalyticity in the level density.
An experimental realization of the model in the presently considered extended form would provide an excellent playground for testing the effects that accompany various kinds of quantum criticality.
Such studies can be based on a suitable extension of the current experiment reported in Ref.\,\cite{Cai21}.

Our main attention has been focused to the ESQPT caused by the $(q,p)\I{=}(0,0)$ stationary point, which is present in the parity-conserving model and the ${\gamma=1}$ form of the parity-violating model. 
This point corresponds to the real vacuum of the field, yielding zero expectation values of the boson creation and annihilation operators (in contrast to the Higgs-like \uvo{vacua} associated with the ground state of the system in the superradiant phases).
Metamorphoses of this stationary point in transitions between various parameter domains define different \uvo{vacuum phases} of the system.
While in the normal phase the vacuum represents a stable minimum of the Hamiltonian, in various superradiant phases it can be either a local minimum, a saddle point, or a local maximum.
Local stability properties of the vacuum determine the type of the corresponding ESQPT singularity and are essential for the stabilization properties of the vacuum-involving factorized qubit-field state in quantum evolution.

In particular, the saddle point of the classical-limit Hamiltonian, which corresponds to the vacuum state for parity-conserving interactions in the first superradiant phase, induces a stabilization of the qubit state but makes the field state depart from the initial form.
On the other hand, the local maximum of the Hamiltonian, which characterizes the vacuum state for ultrastrong parity-conserving interactions in the second superradiant phase, leads to stabilization of both qubit and field states.
A partial stabilization of this kind occurs even in presence of parity-violating interactions and is universal if the parity-conserving interactions dominate over the parity-violating ones.
The degree of the qubit stabilization effect increases in an approximately algebraic way with $R$ and becomes ultimate in the ${R\to\infty}$ limit.

The stabilization effect and its variations in the parameter space can be used as a measurable signature of the corresponding types of ESQPTs in various experimental realizations of the extended Rabi Hamiltonian.
Very similar effects (including the difference between the first and second superradiant phases) were numerically observed also in the Dicke atom-field model with ${N\I{\gg}1}$ \cite{Klo17,Klo18}.
Related ESQPT-induced stabilization phenomena were reported also in some other types of strongly interacting many-body model systems \cite{San15,San16}.
It needs to be mentioned, however, that for some quench protocols the presence of a stationary point in the semiclassical energy landscape does not lead to an initial-state stabilization, but to a~nearly opposite effect \cite{Per11,Klo18,Klo21}.
We stress that these and related ESQPT-induced dynamical phenomena may play important roles in quantum-state engineering protocols based on diverse model platforms with limited numbers of effective degrees of freedom.

\section*{Acknowledgments}

The authors are grateful to Michal Macek, Michal Kloc and Luk{\'a}{\v s} Slodi{\v c}ka for important discussions.
P.S. and P.C. acknowledge support of the Czech Science Foundation (grant no.\,20-09998S) and the Charles University (project UNCE/SCI/013). 
R.F. acknowledges support of the Czech Science Foundation (grant no.\,20-16577S).


\begin{thebibliography}{99}
%
\bibitem{Qcomp}  M.A. Nielsen and I.L. Chuang, {\it Quantum Computation and Quantum Information} (Cambridge University Press, Cambridge, UK, 2010).
\bibitem{Geo14} I.M. Georgescu, S. Ashhab, and F. Nori, Rev. Mod. Phys. {\bf 86}, 153 (2014).
\bibitem{Har96} S. Haroche and J.‐M. Raimond, Physics Today 49, 8, 51 (1996).
\bibitem{Qerr} D.A. Lidar and T.A. Brun (editors), {\it Quantum Error Correction} (Cambridge University Press, Cambridge, UK, 2013).
%
\bibitem{Rab36} I.I. Rabi, Phys. Rev. {\bf 49}, 324 (1936); {\bf 51}, 652 (1937).
\bibitem{Dic54} R.H. Dicke, Phys. Rev. {\bf 93}, 99 (1954).
\bibitem{Jay63} E.T. Jaynes and F.W. Cummings, Proc. IEEE {\bf 51}, 89 (1963).
\bibitem{Hep73} K. Hepp and E. H. Lieb, Ann. Phys. (NY) {\bf 76}, 360 (1973).
\bibitem{Wan73} Y.K. Wang and F.T. Hioe, Phys. Rev. A {\bf 7}, 831 (1973).
\bibitem{Rza75} K. Rzazewski, K. Wodkiewicz and W. Zakowicz, Phys. Rev. Lett. {\bf 35}, 432 (1975).
\bibitem{Ema03} C. Emary and T. Brandes, Phys. Rev. Lett. {\bf 90}, 044101 (2003); Phys. Rev. E {\bf 67}, 066203 (2003).
\bibitem{Bra05} T Brandes, Phys. Rep. {\bf 408}, 315 (2005).
\bibitem{Bau11} K. Baumann, R. Mottl, F. Brennecke and T. Esslinger, Phys. Rev. Lett. {\bf 107}, 140402 (2011).
\bibitem{Klo17} M. Kloc, P. Str{\'a}nsk{\'y}, P. Cejnar, Ann. Phys. (NY) {\bf 382}, 85 (2017).
\bibitem{Bas17} M.A. Bastarrachea-Magnani, A. Rela{\~n}o, S. Lerma-Hern{\'a}ndez, B. L{\'o}pez-del-Carpio, J. Ch{\'a}vez-Carlos, and J.G. Hirsch, J. Phys. A: Math. Theor. {\bf 50}, 144002 (2017).
%
\bibitem{Bra11} D. Braak, Phys. Rev. Lett. {\bf 107}, 100401 (2011).
%
\bibitem{For10} P. Forn-D{\'\i}az, J. Lisenfeld, D. Marcos, J. J. Garc{\'\i}a-Ripoll, E. Solano, C. J. P. M. Harmans, and J. E. Mooij, Phys. Rev. Lett. {\bf 105}, 237001 (2010).
\bibitem{Nie10} T. Niemczyk,  F. Deppe, H. Huebl, E. P. Menzel, F. Hocke, M. J. Schwarz, J. J. Garcia-Ripoll, D. Zueco, T. H{\"u}mmer, E. Solano, A. Marx, and R. Gross, Nat. Phys. {\bf 6}, 772 (2010).
\bibitem{Bra17} J. Braum{\"u}ller, M. Marthaler, A. Schneider, A. Stehli, H. Rotzinger, M. Weides, and A.V. Ustinov, Nat. Commun. {\bf 8}, 779 (2017).
\bibitem{For17} P. Forn-D{\'\i}az, J.J. Garc{\'\i}a-Ripoll, B. Peropadre, J.-L. Orgiazzi, M.A. Yurtalan, R. Belyansky, C.M. Wilson, and A. Lupascu, Nat. Phys. {\bf 13}, 39 (2017).
\bibitem{Yos17} F. Yoshihara, T. Fuse, S. Ashhab, K. Kakuyanagi, S. Saito, and K. Semba, Nat. Phys. {\bf 13}, 44 (2017).
\bibitem{Lan17} N.K. Langford, R. Sagastizabal, M. Kounalakis, C. Dickel, A. Bruno, F. Luthi, D. J. Thoen, A. Endo, and L. DiCarlo, Nat. Commun. {\bf 8}, 1715 (2017).
\bibitem{Cre12} A. Crespi, S. Longhi, and R. Osellame, Phys. Rev. Lett. {\bf 108}, 163601 (2012).
\bibitem{Tod09} Y. Todorov, A. M. Andrews, I. Sagnes, R. Colombelli, P. Klang, G. Strasser, and C. Sirtori, Phys. Rev. Lett. {\bf 102}, 186402 (2009).
\bibitem{Gun09} G. G{\"u}nter, A.A. Anappara, J. Hees, A. Sell, G. Biasiol, L. Sorba, S. De Liberato, C. Ciuti, A. Tredicucci, A. Leitenstorfer, and R. Huber, Nature {\bf 458}, 178 (2009).
\bibitem{Lv18} D. Lv, S. An, Z. Liu, J.-N. Zhang, J.S. Pedernales, L. Lamata, E. Solano, and K. Kim, Phys. Rev. X {\bf 8}, 021027 (2018).
\bibitem{Lo15} H.-Y. Lo,  D. Kienzler, L. de Clercq, M. Marinelli, V. Negnevitsky, B.C. Keitch, and J.P. Home, Nature {\bf 521}, 336 (2015).
\bibitem{Kie16} D. Kienzler, C. Fl{\"u}hmann, V. Negnevitsky, H.-Y. Lo, M. Marinelli, D. Nadlinger, and J.P. Home, Phys. Rev. Lett. {\bf 116}, 140402 (2016).
\bibitem{Flu18} C. Fl{\"u}hmann, V. Negnevitsky, M. Marinelli, and J.P. Home, Phys. Rev. X {\bf 8}, 021001 (2018).
\bibitem{Flu19} C. Fl{\"u}hmann,  T. L. Nguyen, M. Marinelli, V. Negnevitsky, K. Mehta, and J.P. Home, Nature {\bf 566}, 513 (2019).  
\bibitem {Par17} K. Park, P. Marek, and R. Filip, Sci. Rep. {\bf 7}, 11536 (2017).
\bibitem{Par18} K. Park, P. Marek, and R. Filip, New J. Phys. {\bf 20}, 053022 (2018).
\bibitem{Sta20} R. Stassi, M. Cirio, and F. Nori, Quantum Inf. {\bf 6}, 67 (2020).
\bibitem{Has21a} J. Hastrup,  K. Park, J.B. Brask, R. Filip, and U.L. Andersen,  Quantum Inf. {\bf 7}, 17 (2021).
\bibitem{Has21b} J. Hastrup,  K. Park, R. Filip, and U.L. Andersen, Phys. Rev. Lett. {\bf 126}, 153602 (2021).
\bibitem{Che21} Y.-H. Chen, W. Qin, X. Wang, A. Miranowicz, and F. Nori, Phys. Rev. Lett. {\bf 126}, 023602 (2021).
\bibitem{Ste19} O. Di Stefano, A. Settineri, V. Macr{\`\i}, L. Garziano, R. Stassi, S. Savasta, and F. Nori, Nat. Phys. {\bf 15}, 803 (2019).
\bibitem{Gar16} L. Garziano, V. Macr{\`\i}, R. Stassi, O. Di Stefano, F. Nori, and S. Savasta, Phys. Rev. Lett. {\bf 117}, 043601 (2016). 
\bibitem{Gar20} L. Garziano, A. Ridolfo, A. Miranowicz, G. Falci, S. Savasta, and F. Nori, Sci. Rep. {\bf 10}, 21660 (2020).
\bibitem{Koc17a} A.F. Kockum,  A. Miranowicz, V. Macr{\`\i}, S. Savasta, and F. Nori,  Phys. Rev. A {\bf 95}, 063849 (2017).
\bibitem{Koc17b} A.F. Kockum, V. Macr{\`\i}, L. Garziano, S. Savasta, and F. Nori,  Sci. Rep. {\bf 7}, 5313 (2017).
\bibitem{Sta17} R. Stassi, V. Macr{\`\i}, A.F. Kockum, O. Di Stefano, A. Miranowicz, S. Savasta, and F. Nori,  Phys. Rev. A {\bf 96}, 023818 (2017).
\bibitem{Gua18} G. Guarnieri, M. Kol{\' a}{\v r}, and R. Filip, Phys. Rev. Lett. {\bf 121}, 070401 (2018).
\bibitem{Pur20} A. Purkayastha, G. Guarnieri, M.T. Mitchison, R. Filip, and J. Goold, Quantum Inf. {\bf 6}, 27 (2020).
%
\bibitem{Ash13} S. Ashhab, Phys. Rev. A {\bf 87}, 013826 (2013).
\bibitem{Hwa15} M.-J. Hwang, R. Puebla, and M. B. Plenio, Phys. Rev. Lett. {\bf 115}, 180404 (2015).
\bibitem{Pue16a} R. Puebla, J. Casanova, and M.B. Plenio, New J. Phys. {\bf 18}, 113039 (2016).
\bibitem{Pue16} R. Puebla, M.-J. Hwang, and M. B. Plenio, Phys. Rev. A {\bf 94}, 023835 (2016).
\bibitem{Pue18} R. Puebla, {\it Equilibrium and Nonequilibrium Aspects of Phase Transitions in Quantum Physics} (Springer, Cham, 2018).
\bibitem{Pue20} R.~Puebla, A.~Smirne, S.F.~Huelga, and M.B. Plenio, Phys. Rev. Lett. {\bf 124}, 230602 (2020).
\bibitem{Pue20b} R. Puebla, Phys. Rev. B {\bf 102}, 220302(R) (2020).
\bibitem{Cai21} M.-L. Cai,  Z.-D. Liu, W.-D. Zhao, Y.-K. Wu, Q.-X. Mei, Y. Jiang, L. He, X. Zhang, Z.-C. Zhou, and L.-M. Duan, Nat. Commun. {\bf 12}, 1126 (2021).
%
\bibitem{Car10} L.\,D.~Carr (editor), {\it Understanding Quantum Phase Transitions} (Taylor \& Francis, Boca Raton, 2010). 
%
\bibitem{Cej06} P. Cejnar, M. Macek, S. Heinze, J. Jolie, and J. Dobe{\v s}, J. Phys. A: Math. Gen. {\bf 39}, L515 (2006).
\bibitem{Cap08} M.A.~Caprio, P.~Cejnar, and F.~Iachello, Ann. Phys. (NY) {\bf 323}, 1106 (2008).
\bibitem{Cej08} P. Cejnar and P. Str{\'a}nsk{\'y}, Phys. Rev. E {\bf 78}, 031130 (2008).
\bibitem{Per09} P. P{\'e}rez-Fern{\'a}ndez, A. Rela{\~n}o, J.M. Arias, J. Dukelsky, and J.E. Garc{\'\i}a-Ramos, Phys. Rev.
A {\bf 80}, 032111 (2009).
\bibitem{Per11}P. P{\'e}rez-Fern{\'a}ndez, P. Cejnar, J.M. Arias, J. Dukelsky, J.E. Garc{\'\i}a-Ramos, and A. Rela{\~n}o, Phys. Rev. A {\bf 83}  033802 (2011).
\bibitem{Lar13} D.~Larese, F.~P{\'e}rez-Bernal, and F.~Iachello, J. Mol. Struct. {\bf 1051}, 310 (2013).
\bibitem{Bra13} T.~Brandes, Phys. Rev. E {\bf 88}, 032133 (2013).
\bibitem{Die13} B.~Dietz, F.~Iachello, M.~Miski-Oglu, N.~Pietralla, A.~Richter, L.von~Smekal, and J.~Wambach,  Phys. Rev. B {\bf 88}, 104101 (2013).
\bibitem{Bas14a} M.A.~Bastarrachea-Magnani, S.~Lerma-Hern{\'a}ndez, and J.G. Hirsch, Phys. Rev. A {\bf 89}, 032101,  032102 (2014).
\bibitem{Bas14} V.M.~Bastidas, P.~P{\'e}rez-Fern{\' a}ndez, M.~Vogl, and T.~Brandes, Phys. Rev. Lett. {\bf 112}, 140408 (2014).
\bibitem{San15} L.F. Santos and F. P{\'e}rez-Bernal, Phys. Rev. A {\bf 92}, 050101(R) (2015). 
\bibitem{San16} L.F. Santos, M. T{\'a}vora and F. P{\'e}rez-Bernal, Phys. Rev. A {\bf 94}, 012113 (2016).
\bibitem{Str16} P.~Str{\'a}nsk{\'y} and P. Cejnar, Phys. Lett. A {\bf 380}, 2637 (2016). 
\bibitem{Rel16} A.~Rela{\~n}o, C. Esebbag, and J. Dukelsky, Phys. Rev. E {\bf 94}, 052110 (2016).
\bibitem{Klo18} M. Kloc, P. Str{\'a}nsk{\'y}, and P. Cejnar, Phys. Rev. A {\bf 98}, 013836 (2018).
\bibitem{Kha19} J.~Khalouf-Rivera, M.~Carvajal, L.F.~Santos, and F.~P{\'e}rez-Bernal, J. Phys. Chem. A {\bf 123}, 9544 (2019).
\bibitem{Mac19} M.~Macek, P.~Str{\'a}nsk{\'y}, A.~Leviatan, and P.~Cejnar, Phys. Rev. C {\bf 99}, 064323 (2019).
\bibitem{Wan19} Q. Wang and F. P{\'e}rez-Bernal,  Phys. Rev. A {\bf 100}, 022118 (2019).
\bibitem{Hum19} Q. Hummel, B. Geiger, J.D. Urbina, and K. Richter, Phys. Rev. Lett. {\bf 123}, 160401 (2019).
\bibitem{Tia20} T.~Tian, H.-X.~Yang, L.-Y.~Qiu, H.-Y.~Liang, Y.-B.~Yang, Y.~Xu, and L.-M.~Duan, Phys. Rev. Lett. {\bf 124}, 043001 (2020).
\bibitem{Mum20} J. Mumford, W. Kirkby, and D.H.J. O'Dell, J. Phys. B: At. Mol. Opt. Phys. {\bf 53}, 145301 (2020).
\bibitem{Str20} P.~Str{\'a}nsk{\'y}, M~{\v S}indelka, M. Kloc, and P.~Cejnar, Phys. Rev. Lett. {\bf 125}, 020401 (2020); P.~Str{\'a}nsk{\'y}, M~{\v S}indelka, and P.~Cejnar, Phys. Rev. A {\bf 103}, 062207 (2021).
\bibitem{Fel21} P.~Feldmann, C.~Klempt, A.~Smerzi, L.~Santos, and M~Gessner, Phys. Rev. Lett. {\bf 126}, 230602 (2021).
\bibitem{Klo21} M. Kloc, D. {\v S}imsa, F. Han{\' a}k, P.R. Kapr{\'a}lov{\'a}-{\v Z}{\v d}{\'a}nsk{\'a}, P. Str{\'a}nsk{\'y}, and P. Cejnar, Phys. Rev. A {\bf 103} 032213 (2021).
\bibitem{Cab21} J. Cabedo, J. Claramunt, A. Celi, arXiv: 2101.08253 [cond-mat.quant-gas] (2021).
\bibitem{Cej21} P. Cejnar, P. Str{\'a}nsk{\'y}, M. Macek, and M. Kloc, J. Phys. A: Math. Theo. {\bf 54}, 133001 (2021).
\bibitem{Kra18} S. Kr{\"a}mer, D. Plankensteiner, L. Ostermann, and H. Ritsch, Comp. Phys. Comm. {\bf 227}, 109 (2018).
\end{thebibliography}
\end{document}